\documentclass[preprint,amsmath,amssymb]{revtex4}
\usepackage{dcolumn}
\usepackage{bm}
\usepackage{graphicx}
\usepackage{subfigure}
\usepackage{color}

\newcommand{\be}{\begin{eqnarray}}
\newcommand{\ee}{\end{eqnarray}}

\begin{document}

\title{Generalized free energy landscapes of the charged Gauss-Bonnet AdS black holes in diverse dimensions }

\author{Ran Li$^{a}$}
\thanks{liran@wiucas.ac.cn}

\author{Jin Wang$^{b}$}
\thanks{Corresponding author, jin.wang.1@stonybrook.edu}

\affiliation{$^a$ Center for Theoretical Interdisciplinary Sciences, Wenzhou Institute, University of Chinese Academy of Sciences, Wenzhou, Zhejiang 325001, China}

\affiliation{$^b$ Department of Chemistry and Department of Physics and Astronomy, State University of New York at Stony Brook, Stony Brook, New York 11794, USA}

\begin{abstract}
The present study focuses on analyzing the generalized free energy function of the $D$-dimensional charged Gauss-Bonnet AdS black holes. We examine the fluctuating black holes that are in contact with thermal baths at an arbitrary ensemble temperature, resulting in the corresponding Euclidean geometry with a conical singularity at the event horizon. By properly regularizing the conical singularity, we have derived the generalized free energy of the fluctuating black holes in the canonical ensemble using the Euclidean gravitational path integral approach. We demonstrate that the derived generalized free energy is consistent with the definition from thermodynamic perspective. Then we explore the free energy landscapes of the charge Gauss-Bonnet black holes in diverse spacetime dimensions and examine the corresponding thermodynamics of the black hole phase transition. Finally, we discuss the generalized free energy landscape of the fluctuating black holes in grand canonical ensemble.   
\end{abstract}

\maketitle

\section{introduction}

Since the discovery of the black hole event horizon radiating in a thermal spectrum with the temperature proportional to its surface gravity \cite{Hawking:1975vcx}, studying the phase transition of black holes from the thermodynamic perspective has become an important topic in the intersection among general relativity, thermodynamics, and statistical physics. In recent years, by treating the cosmological constant as the thermodynamic pressure and introducing the concept of thermodynamic volume of black hole \cite{Kastor:2009wy,Dolan:2010ha,Dolan:2011xt,Dolan:2011jm,Cvetic:2010jb}, it is revealed that the charged black holes in AdS space exhibit the analogous behavior to the Van der Waals fluids \cite{Kubiznak:2012wp}. This observation stimulates a series of new discoveries, such as reentrant phase transitions \cite{Gunasekaran:2012dq,Altamirano:2013ane}, triple critical points \cite{Altamirano:2013uqa,Wei:2014hba} and multiple critical points \cite{Tavakoli:2022kmo,Wu:2022plw}, holographic heat engine \cite{Johnson:2014yja}, Ruppeiner geometry and microstructure of black holes \cite{Wei:2019uqg}, and topological classification of black holes \cite{Wei:2022dzw}, et. al.

With in this context, the free energy landscape is shown to be a valuable concept and tool for examining the thermodynamics and the kinetics of black hole and its phase transitions \cite{Li:2020khm,Li:2020nsy}. It is assumed that during the transition process from one local stable state to another, the system can pass through a series of intermediate states. These intermediate states are formed due to the thermal fluctuations. For this reason, they are also called the fluctuating black holes \cite{Li:2021vdp}. To properly describe the fluctuating black holes, one can introduce the order parameter and the generalized free energy function. The generalized free energy of the fluctuating black hole was initially defined by using the thermodynamic relations. In \cite{Li:2022oup}, it is demonstrated that the generalized free energy can be derived from gravitational action by using the Euclidean path integral approach.

In this way, one can illustrate the topography of the generalized free energy function graphically, which is also known as the free energy landscape \cite{FSW,FW}. The free energy provides the weight or the probability of each state in the system. The free energy landscape can be used to connect one state to another. If only one order parameter is introduced to characterize the black hole's microscopic state, the free energy landscape provides an intuitive representation of the one-dimensional topography of the generalized free energy function. The free energy landscape gives a global quantification and characterization of the system, for example, the global stability. It is well known that the free energy landscape can provide not only the topography that determines the thermodynamic stability of the system but also the thermal potential that drives the dynamical process and the phase transition \cite{JW,JWRMP}.

Due to the existence of thermal fluctuations, the dynamical process of black hole state transition and phase transition is then described stochastically by using Langevin equation that gives the time evolution of the black hole order parameter and the Fokker-Planck equation that gives the time evolution of the probability distribution of the black hole states. Under these assumptions, it is shown that the kinetics of the black hole state switching and phase transition is mainly determined by the barrier height of the free energy landscape \cite{Li:2022ylz,Li:2022yti,Wei:2020rcd,Li:2020spm,Lan:2021crt,Li:2021zep,Yang:2021ljn,Mo:2021jff,Kumara:2021hlt,Li:2021tpu,Liu:2021lmr,Xu:2021usl,Du:2021cxs,Dai:2022mko,Luo:2022gss,Xu:2022jyp,Ali:2023wkq}. A very interesting example shows that at the triple point of the six-dimensional Gauss-Bonnet gravity, the probability distribution of the black hole states exhibits an oscillating behaviour in the dynamical transition process \cite{Wei:2021bwy}.

The one-dimensional free energy landscape has recently been extended to two dimensions \cite{Li:2023ppc}. In the case of the five dimensional charged Gauss-Bonnet AdS black hole in the grand canonical ensemble \cite{Cai:2013qga,Zou:2014mha}, two order parameters, black hole radius and charge, are introduced to characterize the microscopic state of the charged Gauss-Bonnet AdS black hole. The generalized free energy as a function of black hole radius and charge is then defined in terms of the thermodynamic relation in the grand canonical ensemble. However, the derivation of the generalized free energy function of the charged Gauss-Bonnet AdS black hole from the gravitational path integral approach was not yet been carried out. In this work, we address this issue and aim to derive of the generalized free energy function for Gauss-Bonnet gravity in $D-$dimensional spacetime.

In the derivation of gravitational action of the fluctuating black holes, there are two issues that should be carefully managed \cite{Li:2022oup}. One is to handle the conical singularity in the Euclidean geometry of the fluctuating black hole. Another one is to eliminate the bulk divergence due to the volume infinity of the AdS space. For the first one, we employ the smooth regularization method to compute the conical singularity's contribution to the action \cite{Fursaev:1995ef,Solodukhin:1994yz,Fursaev:1994te,Solodukhin:2011gn}, and for the second one, we use the subtraction trick to get the finite part of the action for the AdS black holes \cite{Gibbons:2004ai}. After derivation of the generalized free energy function from the gravitational action, we discuss the free energy landscapes for the Gauss-Bonnet gravity in $D=5$, $D=6$, and $D\ge7$. It is shown that for $D=5$ and $D\ge7$, there is only one critical point on the phase diagram representing the endpoint of the coexisting curve of the small and the large black holes. For $D=6$, there exist two cases, the first one is that there is only one critical points and the second one is that there are two critical points and one triple point on the phase diagram. The second case indicates that at the phase transition point, the shape of the free energy landscape has three wells. For the first case, although the shape of the landscape is double well at the phase transition point, it still has three wells at a specific temperature range. Based on the topography of the landscape, we discussed the thermodynamics of the phase transition. We also study the generalized free energy function for the Gauss-Bonnet black holes in the grand canonical ensemble, which is considered as the function of two order parameters. The corresponding two dimensional free energy landscapes are also illustrated.

This paper is arranged as follows. In Sec.\ref{SecII}, we will briefly review the basic facts about the $D$-dimensional Gauss-Bonnet AdS black holes and introduce the Euclidean geometry of the fluctuating Gauss-Bonnet AdS black hole. In Sec.\ref{SecIII}, we calculate the partition function of Gauss-Bonnet gravity in canonical ensemble and derive the generalized free energy function for the fluctuating black holes. In Sec.\ref{SecIV}, by treating the generalized free energy as the function of the black hole radius, we show the free energy landscapes in diverse dimensions. The thermodynamic stabilities and the corresponding phase diagrams are discussed in detail. In Sec.\ref{SecV}, we give the generalized free energy of the fluctuating black holes in grand canonical ensemble. At last, the conclusion and the discussions are presented in Sec.\ref{SecVI}.

\section{Euclidean geometry of the fluctuating Gauss-Bonnet black hole}

In the dynamical process of the black hole state switching and phase transition, due to the influence of the thermal fluctuations, one local stable black hole state can pass through a series of intermediate black hole states to reach another local stable state. These intermediate black hole states are formed due to the thermal fluctuations, which are named as the fluctuating black holes. The fluctuating black holes are considered to be the intermediate states during the black hole state switching and phase transition process. In this section, we consider the Euclidean geometry of the fluctuating Gauss-Bonnet black hole.

\subsection{$D-$dimensional Gauss-Bonnet black holes}

We start with the metric of the general spherically symmetric $D$ dimensional charged Gauss-Bonnet black hole in AdS space, which is given by \cite{Boulware:1985wk,Cai:2001dz,Wiltshire:1985us}
\begin{eqnarray}
ds^2=- f(r) dt^2+\frac{1}{f(r)}dr^2+r^2 d\Omega^2_{D-2}\;,
\end{eqnarray}
where $d\Omega^2_{D-2}$ is the metric of the unit $(D-2)$-dimensional sphere. The black factor $f(r)$ is given by
\begin{eqnarray}
 f(r)=1+\frac{r^2}{2\alpha} \left[ 1-
 \sqrt{1+4\alpha\left(\frac{\omega}{r^{D-1}}-\frac{q^2}{r^{2D-4}}-\frac{1}{L^2}\right)}\right]\;,
\end{eqnarray}
where $\omega$ and $q$ are related to the mass and charge of the black hole, $\alpha$ is the Gauss-Bonnet coupling constant, and $L$ is the AdS radius related to the cosmological constant $\Lambda$ by the equality $\Lambda=-\frac{(D-1)(D-2)}{2L^2}$. It should be noted that, in order to have a well-defined vacuum in the theory, the effective Gauss-Bonnet coefficient $\alpha$ should satisfy the constraint \cite{Cai:2013qga,Zou:2014mha}
\begin{eqnarray}
0\leq \frac{4 \alpha}{L^2}\leq 1\;. 
\end{eqnarray}

The $U(1)$ gauge field or the electromagnetic gauge field is given by 
\begin{eqnarray}
    A_t(r)=-\frac{1}{\sqrt{8\pi}}\sqrt{\frac{D-2}{D-3}}\frac{q}{r^{D-3}}+\Phi_H\;,
\end{eqnarray}
where $\Phi_H=\frac{1}{\sqrt{8\pi}}\sqrt{\frac{D-2}{D-3}}\frac{q}{r_h^{D-3}}$ is introduced to guarantee the regularity of electromagnetic gauge field at the horizon. It is easy to check that the metric together with the electromagnetic field given above solves the equations of motion derived from varying the Einstein-Gauss-Bonnet action.

The black hole horizon is determined by the equation $f(r_h)=0$. The black hole's Hawking temperature is then given by \cite{Hawking:1975vcx}
\begin{eqnarray}\label{Hawk_temp}
    T_H=\frac{1}{4\pi} \left.f'(r)\right|_{r=r_h}=\frac{(D-3)r_h^2+(D-1)r_h^4/L^2+(D-5)\alpha-(D-3)q^2/r_h^{2(D-4)}}{4\pi r_h \left(r_h^2+2\alpha\right)}
\end{eqnarray}
The mass, charge and Bekenstein-Hawking entropy of the Gauss-Bonnet AdS black hole are given by \cite{Myers:1988ze,Cvetic:2001bk,Haroon:2020vpr,Marks:2021fpe}
\begin{eqnarray}\label{M_Q_S}
    M&=&\frac{(D-2)\Omega_{D-2}}{16\pi}\omega= \frac{(D-2)\Omega_{D-2}}{16\pi} \left(r_h^{D-3}+r_h^{D-1}/L^2+\alpha r_h^{D-5}+q^2/r_h^{D-3}\right)
 \nonumber\\
    Q&=&\sqrt{\frac{(D-2)(D-3)}{8\pi}}\Omega_{D-2} q\nonumber\\
    S&=& \frac{1}{4}  \Omega_{D-2} r_h^{D-2} \left(1+\frac{2\alpha(D-2)}{(D-4)}\frac{1}{r_h^2}\right)
\end{eqnarray}
where we have expressed these quantities by using the black hole radius $r_h$ and the charge parameter $q$ for latter convenience.

\subsection{Euclidean geometry of the fluctuating Gauss-Bonnet black hole}\label{SecII}

We now consider the case that the Gauss-Bonnet black hole is in contact with a thermal bath at a fixed temperature $T$. This is to say we have introduced the canonical ensemble description of the black holes. The temperature of the thermal bath is just the canonical ensemble's temperature. Because our aim is to study the Euclidean geometry of the fluctuating Gauss-Bonnet black hole generated during the phase transition process under the influence of thermal noises, the intrinsic Hawking temperature of the fluctuating black hole is not necessarily equal to the temperature of the thermal bath in general. This will introduce a conical singularity in the Euclidean geometry of the black hole.

Firstly, we introduce the Euclidean time $\tau=it$ by a Wick rotation in the complex $t$ plane. Then the metric becomes 
\begin{eqnarray}\label{euclidean_metric}
ds^2=f(r) d\tau^2+\frac{1}{f(r)}dr^2+r^2 d\Omega^2_{D-2}\;.
\end{eqnarray}
As we have discussed, the presence of the thermal bath or the environment results in that the periodicity of Euclidean time $\tau$ is determined by the canonical ensemble's parameter by the relation  
\begin{eqnarray}
0\leq \tau \leq \beta\;,
\end{eqnarray}
where $\beta=\frac{1}{T}$ is the inverse of the ensemble temperature $T$. It should be noted that, in this setup, the period $\beta$ of the Euclidean time is unrelated to any parameters of black hole, such as mass, charge, or black hole radius, et. al.

Near the horizon, by introducing the coordinate $\rho$ as
\begin{eqnarray}
\rho=\int\frac{dr}{\sqrt{f(r)}}\simeq \frac{2}{\sqrt{f'(r_h)}}\sqrt{r-r_h}\;,
\end{eqnarray}
the metric of the Euclidean Gauss-Bonnet AdS black hole can be approximated by 
\begin{eqnarray}\label{NH_metric}
ds^2\simeq \rho^2d\left(\frac{2\pi \tau}{\beta_H}\right)^2+d\rho^2+r_h^2d\Omega_{D-2}^2\;,
\end{eqnarray}
where $\beta_H=4\pi/f'(r_h)$ is the inverse Hawking temperature of the Gauss-Bonnet black hole.

\begin{figure}
  \centering
  \includegraphics[width=8cm]{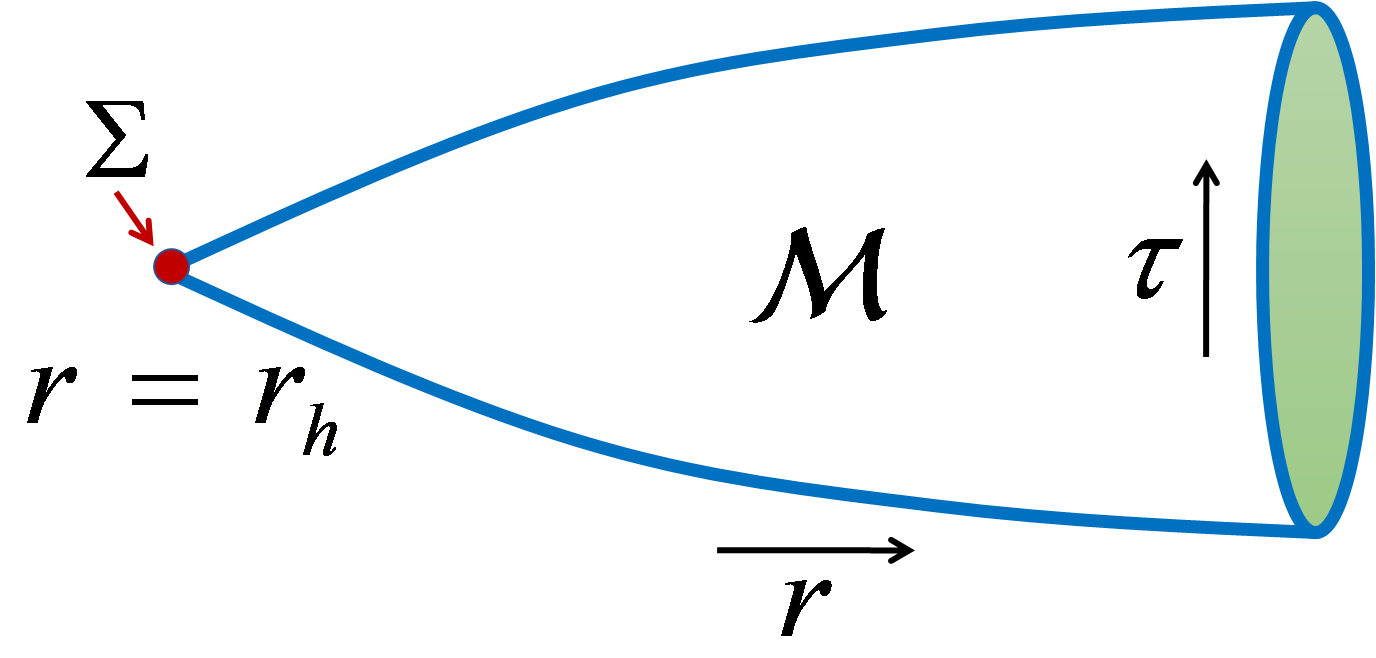}\\
  \caption{An two dimensional illustration of the Euclidean geometry $\mathcal{M}$ that describes the fluctuating Gauss-Bonnet AdS black hole. Every point in this two dimensional surface represents a $(D-2)$ dimensional sphere of radius $r$. The event horizon radius is located at $r=r_h$ and the Euclidean time period is arbitrary value $\beta$. The red point represents the conical singularity $\Sigma$ at the event horizon $r=r_h$. When $\beta$ is equal to the inverse temperature $\beta_H$, the conical singularity disappears and the corresponding Euclidean geometry becomes regular.}
  \label{Euclidean_Geometry}
\end{figure}

We are considering the case that the Euclidean time $\tau$ has an arbitrary period $\beta$. Therefore, near the horizon, the Euclidean geometry of the fluctuating Gauss-Bonnet AdS black hole described by the metric (\ref{NH_metric}) represents the product manifold of a two dimensional cone and a $(D-2)$ dimensional sphere \cite{Fursaev:1995ef,Solodukhin:1994yz,Fursaev:1994te,Solodukhin:2011gn}. The Euclidean geometry of the fluctuating Gauss-Bonnet AdS black hole is depicted in Figure \ref{Euclidean_Geometry}. The manifold is not smooth in the present case. One should note that when the periodicity of the Euclidean time is set to be the inverse Hawking temperature $\beta_H$, the conical singularity $\Sigma$ generates into a two dimensional disk and the corresponding Euclidean geometry becomes regular. From our previous work \cite{Li:2022oup}, we know that when the equilibrium condition $\beta=\beta_H$ is satisfied, the black hole is in locally stable state, otherwise the black hole becomes the fluctuating one on the free energy landscape.

For the fluctuating Gauss-Bonnet AdS black hole, there is an deficit angle $2\pi\left(1-\frac{\beta}{\beta_H}\right)$ to describe the canonical singularity $\Sigma$ at the event horizon $r=r_h$. One can also observe that because of the existence of canonical singularity, at the horizon $r_h$, the Euclidean metric of Gauss-Bonnet AdS black hole cannot satisfy the equations of motion for the Einstein-Gauss-Bonnet gravity. However, in the AdS bulk except the event horizon, the equations of motion is still satisfied because the metric or the geometry is regular there. Therefore, to calculate the gravitational partition function by using the semiclassical approximation method, one just needs to compute the Einstein-Gauss-Bonnet gravitational action on the singular Euclidean manifold as gravitational instanton which is depicted in Figure \ref{Euclidean_Geometry}, although special attentions should be taken on the conical singularity's contribution.

\section{Partition function of the fluctuating Gauss-Bonnet black hole in canonical ensemble} \label{SecIII}

In this section, we calculate the partition function of $D-$dimensional fluctuating Gauss-Bonnet AdS black hole by using the Gibbons-Hawking path integral approach \cite{Gibbons:1976ue}. 

\subsection{Einstein-Gauss-Bonnet action}

In Gibbons-Hawking approach to black hole thermodynamics, the partition function of black hole in the canonical ensemble is given by the gravitational path integral \cite{Gibbons:1976ue}
\begin{eqnarray}\label{partition_func}
Z_{grav}(\beta)=\int D[g] e^{-I_E[g]}\;, 
\end{eqnarray}
where $\beta$ is the period of the Euclidean time representing the inverse temperature of the canonical ensemble, $D[g]$ is the measure of the functional integral, and $I_E[g]$ is the Euclidean gravitational action. This functional integral should be taken on all the Euclidean gravitational configurations that satisfy the given boundary conditions. However, for our purpose, by using the saddle point approximation, we can just evaluate the functional integral over the fluctuating black hole described by the Euclidean metric (\ref{euclidean_metric}). In this way, we can get  
\begin{eqnarray}\label{semi-prox}
Z_{grav}(\beta)\simeq  e^{-I_E[g]} \;, 
\end{eqnarray}
where $g$ is the metric of the fluctuating Gauss-Bonnet black hole given by Eq.(\ref{euclidean_metric}). 

For the Einstein-Gauss-Bonnet gravity coupled with the electromagnetic field, the Euclidean action is given by \cite{Haroon:2020vpr} 
\begin{eqnarray}\label{E_H_action}
I_E=-\frac{1}{16\pi} \int_{\mathcal{M}}\sqrt{g} d^D x  \left( R -2\Lambda+\frac{\alpha}{(D-3)(D-4)}\mathcal{L}_{GB}-4\pi F_{\mu\nu}F^{\mu\nu}\right)
\;,
\end{eqnarray}
where $\mathcal{R}$ is the Ricci scalar curvature, $\Lambda$ is the cosmological constant, and $\alpha$ is the Gauss-Bonnet coupling constant. The Gauss-Bonnet term $\mathcal{L}_{GB}$ is given by 
\begin{eqnarray}
    \mathcal{L}_{GB}=R^2-4 R_{\mu\nu}R^{\mu\nu}+R_{\mu\nu\lambda\rho}R^{\mu\nu\lambda\rho}\;.
\end{eqnarray}

It is well known that there should be boundary terms in the action in order to ensure a well defined variation problem for the equations of motion \cite{Gibbons:1976ue}. In addition, there should be counter terms to cancel the divergence caused by the volume infinity of AdS bulk. However, in the present work, we will utilize the background subtraction trick \cite{Hawking:1982dh,Witten:1998qj} to compute the finite part of the Euclidean gravitational action. In this approach, the boundary terms will be properly cancelled because the black hole's correction to the AdS metric decays very rapidly at the spatial infinity. In the following, we will calculate the Einstein-Gauss-Bonnet action on the singular Euclidean manifold with the conical singularity at the interior boundary.

\subsection{Conical singularity's contribution}

In this subsection, we study the Conical singularity's contribution to the action. Because the curvature tensor is divergent at the conical singularity, a proper regularization scheme is needed. Our strategy is described as follows. Firstly, we regulate the tip of the cone by using a smooth function and calculate the relevant curvature tensors. Then, we take the sharp limit in the end. It turns out that the obtained result for the conical singularity's contribution to the action is independent of the smooth function used in the regularization. 

Our starting point is the Euclidean geometry described by Eq.(\ref{euclidean_metric}). The metric describes the manifold that has a specific product structure of two dimensional cone and $(D-2)-$dimensional sphere. However, without the loss of generality, we specify the metric to be in a more general form as follows \cite{Gregory:2013hja,Burda:2015yfa}
\begin{eqnarray}\label{singular_metric}
    ds^2=d\rho^2+A^2(\rho) d\chi^2+ C^2(\rho) d\Omega_{D-2}^2\;,
\end{eqnarray}
where $(\rho,\chi)$ are the local cylindrical coordinates to parameterize the two dimensional cone. The conical defect or singularity is situated at $\rho\rightarrow 0$ and the transverse space is independent of the local cylindrical coordinates $(\rho,\chi)$ in this limit. In fact, the following calculations are independent of the precise structure of the space transverse to the two dimensional cone, provided $C'(0) =0$. One can define the area of the conical defect to be $C^{D-2}(0)$ times the area given by the metric $\Omega_{D-2}^2$.

The strategy now is to regularize the conical singularity in metric (\ref{singular_metric}) by taking an arbitrary smooth function $A(\rho)$ satisfying the conditions $A'(0) = 1$ and $A'(\epsilon)=(1-\delta)$, where $2\pi\delta$ is the deficit angle. With the metric (\ref{singular_metric}), it is straightforward to calculate that 
\begin{eqnarray}
    R&=&-\frac{2A''(\rho)}{A(\rho)}-2(D-2)\frac{A'(\rho)C'(\rho)}{A(\rho)C(\rho)}+(D-2)(D-3)\frac{\left(1-C'(\rho)^2\right)}{C^2(\rho)}
    \nonumber\\&&
    -2(D-2)\frac{C''(\rho)}{C(\rho)}\;,\nonumber\\
    R^2&=&4\left[\frac{A''(\rho)}{A(\rho)}+(D-2)\frac{A'(\rho)C'(\rho)}{A(\rho)C(\rho)}-(D-2)(D-3)\frac{\left(1-C'(\rho)^2\right)}{4C^2(\rho)}\right.
    \nonumber\\&&\left.+(D-2)\frac{C''(\rho)}{C(\rho)}\right]^2\;,\nonumber\\
    R_{\mu\nu}R^{\mu\nu}&=&\left[(D-2)\frac{A'(\rho)C'(\rho)}{A(\rho)C(\rho)}+\frac{A''(\rho)}{A(\rho)}\right]^2+\left[\frac{A''(\rho)}{A(\rho)}+(D-2)\frac{C''(\rho)}{C(\rho)}\right]^2\nonumber\\
    &&+(D-2)\left[\frac{A'(\rho)C'(\rho)}{A(\rho)C(\rho)}-(D-3)\frac{\left(1-C'(\rho)^2\right)}{C^2(\rho)}+\frac{C''(\rho)}{C(\rho)}\right]^2\;,\nonumber\\
    R_{\mu\nu\lambda\rho}R^{\mu\nu\lambda\rho}&=&
    4\left[\frac{(D-2)(D-3)}{2}\left(\frac{\left(1-C'(\rho)^2\right)}{C^2(\rho)}\right)^2+\left(\frac{A''(\rho)}{A(\rho)}\right)^2\right.
    \nonumber\\&&\left.
    +(D-2)\left(\frac{A'(\rho)C'(\rho)}{A(\rho)C(\rho)}\right)^2+(D-2)\left(\frac{C''(\rho)}{C(\rho)}\right)^2 \right]\;.\nonumber
\end{eqnarray}

After some algebra, one can get 
\begin{eqnarray}
    &&R+\frac{\alpha}{(D-3)(D-4)}\mathcal{L}_{GB}\nonumber\\
    &=&-\frac{2A''(\rho)}{A(\rho)}-2(D-2)\frac{A'(\rho)C'(\rho)}{A(\rho)C(\rho)}-2(D-2)\frac{C''(\rho)}{C(\rho)}\nonumber\\&&-(D-2)(D-3)\frac{\left(1-C'(\rho)^2\right)}{C^2(\rho)}+\frac{8\alpha(D-2)}{(D-4)}\frac{A'(\rho)C'(\rho)C''(\rho)}{A(\rho)C^2(\rho)}\nonumber\\&&
    -\frac{4\alpha(D-2)}{(D-4)}\frac{A''(\rho)}{A(\rho)}\frac{\left(1-C'(\rho)^2\right)}{C^2(\rho)}-4\alpha(D-2)\frac{A'(\rho)C'(\rho)}{A(\rho)C(\rho)} \frac{\left(1-C'(\rho)^2\right)}{C^2(\rho)}\nonumber\\&&
    -4\alpha(D-2) \frac{C''(\rho)}{C(\rho)}\frac{\left(1-C'(\rho)^2\right)}{C^2(\rho)}+\alpha(D-2)(D-5) \left(\frac{\left(1-C'(\rho)^2\right)}{C^2(\rho)}\right)^2  \nonumber\;.
\end{eqnarray}

The idea is now to find out the singular part in the above expression. Because $A'(0) = 1$ and $A'(\epsilon)=(1-\delta)$, $A''(\rho)$ is singular in the vicinity of conical defect because $A''=\mathcal{O}((A'(\epsilon)-A'(0))/\epsilon)$ is unbounded as $\epsilon\rightarrow 0$. In addition, because $C(\rho)$ remains smooth in the vicinity of conical defect, we can expand the function $C(\rho)$ as $C(\rho)=C_0+C_2\rho^2+\cdots$. Therefore, computing the gravitational Lagrangian in the vicinity of $\rho=0$ yields 
\begin{eqnarray}
    R+\frac{\alpha}{(D-3)(D-4)}\mathcal{L}_{GB}&=&-\frac{2A''(\rho)}{A(\rho)}-\frac{4\alpha(D-2)}{(D-4)}\frac{1}{C_0^2}\frac{A''(\rho)}{A(\rho)}
    \nonumber\\&&-4(D-2)\frac{C_2}{C_0}-(D-2)(D-3)\frac{1}{C_0^2}\nonumber\\&&
    -8\alpha(D-2)\frac{C_2}{C_0^3}
    +\alpha(D-2)(D-5)\frac{1}{C_0^4}+\mathcal{O}(\rho)\;.
\end{eqnarray}
It is clear that the Lagrangian is the sum of a regular part (the terms involving $C_0$, $C_2$ and $\mathcal{O}(\rho)$) and the unbounded term $A''/A$.

When computing the integral of the above terms over a small region around the conical singularity $\rho=0$, it is clear that only the first two unbounded terms will contribute to the result, which is given by 
\begin{eqnarray}
  && - \frac{1}{16\pi}\int d^D x \sqrt{g} \left(R+\frac{\alpha}{(D-3)(D-4)}\mathcal{L}_{GB}\right)
  \nonumber\\&=&\frac{1}{8\pi}\Omega_{D-2} r_h^{D-2}\left(A'(0)-A'(\epsilon)\right)\left[1+\frac{2\alpha(D-2)}{D-4}\frac{1}{r_h^2}\right]\nonumber\\&=&
   - \frac{1}{4}\Omega_{D-2} r_h^{D-2}\left(1-\frac{\beta}{\beta_H}\right)\left[1+\frac{2\alpha(D-2)}{D-4}\frac{1}{r_h^2}\right]\;.
\end{eqnarray}
To get this result, we have substituted $C_0=r_h$ and the deficit angle $\delta=2\pi\left(1-\frac{\beta}{\beta_H}\right)$ into the above equation. The final result is independent of the selection of the regularization function $A(\rho)$ as promised. This is our main result for the conical singularity's contribution to the action. Using the expression for black hole entropy in Eq.(\ref{M_Q_S}), it can also be rewritten as $-\left(1-\frac{\beta}{\beta_H}\right)S$. It is shown that this result is proportional to the deficit angle and the entropy of the Gauss-Bonnet black hole.

\subsection{Bulk's contribution} 

Now, we consider the bulk's contribution to the gravitational action. Since the metric (\ref{euclidean_metric}) is static, the time integration gives rise to the period $\beta$ of the Euclidean time. The integration over the transverse space gives rise to the volume of the $(D-2)$-dimensional sphere. The integration on the radial direction requires a little algebra. It can be shown that the bulk action is given by a closed form   
\begin{eqnarray}
&&-\frac{1}{16\pi} \int_{\mathcal{M}/\Sigma}\sqrt{g} d^D x  \left( R -2\Lambda+\frac{\alpha}{(D-3)(D-4)}\mathcal{L}_{GB}\right)\nonumber\\
&=&
-\frac{1}{16\pi} \beta \Omega_{D-2} r^{D-5} \left[
(D-2)\left(r^2+\frac{r^4}{L^2}+\alpha\left(1-\delta_{D,5}\right)-\left(r^2+2\alpha-\alpha f(r)\right) f(r)\right)\right.\nonumber\\&&
\left.-r \left(r^2+\frac{2\alpha(D-2)}{(D-4)}\left(1-f(r)\right)\right)f'(r) \right]\;.
\end{eqnarray}
Note that the integrand in the action is just a total derivative of $r$. This expression gives the indefinite integral of the action. It is divergent when evaluating this expression at the asymptotic AdS spatial infinity. In order to regularize the bulk action, one can employ the background subtracting method. The procedure is firstly to terminate the $r$ integral at a cutoff boundary $r=r_0$, then to subtract off the action of the pure AdS space, and finally to take the limit of $r_0\rightarrow +\infty$ to obtain the finite part of the bulk action \cite{Hawking:1982dh,Witten:1998qj,Gibbons:2004ai}. 

One can set $\omega=0$ and $q=0$ to get the background AdS metric. The factor $f_0(r)$ of the background AdS metric is then given by 
\begin{eqnarray}
    f_0(r)=1+\frac{r^2}{2\alpha} \left[ 1-
 \sqrt{1-\frac{4\alpha}{L^2}}\right]\;.
\end{eqnarray}
In the subtracting procedure, we have to match the Gauss-Bonnet AdS black hole metric with the background AdS metric at the cutoff boundary $r=r_0$. Thus the time coordinate $\tau_0$ of the background AdS space should related to the time coordinate $\tau$ of the Gauss-Bonnet AdS metric by the following equation 
\begin{eqnarray}
f(r_0) d\tau^2= f_0(r_0) d\tau_0^2\;.
\end{eqnarray}
This in turn gives the time periods in the action integrals that are related by the relation 
\begin{eqnarray}
\beta_0=\frac{f(r_0)^{1/2}}{f_0(r_0)^{1/2}} \beta
=1-\frac{\omega L^2}{4}\left(1+\frac{1}{\sqrt{1-\frac{4\alpha}{L^2}}}\right)\frac{1}{r_0^{D-1}}+\mathcal{O}\left(\frac{1}{r_0^{D+1}}\right)\;.
\end{eqnarray}
Thus, the bulk action of the Gauss-Bonnet AdS black hole with the subtraction of the background AdS action is then given by
\begin{eqnarray}
I_{\mathcal{M}/\Sigma}&=&\frac{1}{16\pi}(D-2)\beta\Omega_{D-2} r_h^{D-5}\left(r_h^2+\frac{r_h^4}{L^2}+\alpha\right)\nonumber\\ &&-\frac{1}{4}\frac{\beta}{\beta_H} \Omega_{D-2} r_h^{D-2}\left(1+\frac{2\alpha(D-2)}{(D-4)}\frac{1}{r_h^2}\right)\;,
\end{eqnarray}
where the limit $r_0\rightarrow \infty$ is taken. It can also be written as 
\begin{eqnarray}
I_{\mathcal{M}/\Sigma}=\beta\left(M-\frac{1}{2}Q\Phi_H\right)-\frac{\beta}{\beta_H}S\;.
\end{eqnarray}

\subsection{Electromagnetic field's contribution}

In this subsection, we deal with the contribution to the action from the electromagnetic field.

Because we consider the canonical ensemble with the fixed temperature and treat the electric charge $Q$ as a fixed parameter, we have to add a boundary term to guarantee the fixed charge $Q$ as a boundary condition at infinity. The appropriate action for the electromagnetic field with the boundary term is given by \cite{Caldarelli:1999xj}
\begin{eqnarray}
    I_{EM}&=&\frac{1}{4}\int_{\mathcal{M}} d^D x \sqrt{g} F_{\mu\nu}F^{\mu\nu}-\int_{\partial\mathcal{M}} d^{D-1} x \sqrt{\gamma} n_{\mu}A_{\nu} F^{\mu\nu} \;,
\end{eqnarray}
where $\gamma$ is the induced metric of the boundary $\partial\mathcal{M}$ and $n^{\mu}$ is the outward pointing unit normal vector of the boundary. The variation of the action $I_{EM}$ with respect to the electromagnetic field gives 
\begin{eqnarray}
\delta I_{EM}= -\int_{\mathcal{M}} d^D x \sqrt{g} \left(\nabla_{\mu} F^{\mu\nu} \right) \delta A_{\nu}- \int_{\partial\mathcal{M}} d^{D-1} x \sqrt{\gamma}  A_{\nu} \delta\left(n_{\mu} F^{\mu\nu}\right)\;.
\end{eqnarray}
It is clear that the bulk term gives rise to the equations of motion for electromagnetic field and the surface term yields $\delta\left(n_{\mu} F^{\mu\nu}\right)=0$ as the boundary condition for fixed charge at the AdS infinity. 

By using the stokes theorem, one can transform the bulk integral into the surface integral. The action is then given by 
\begin{eqnarray}
I_{EM}=-\frac{1}{2} \int_{\partial\mathcal{M}} d^{D-1} x \sqrt{\gamma}   n_{\mu}A_{\nu} F^{\mu\nu}\;.
\end{eqnarray}
In Euclidean space, the nonvanishing component of the electromagnetic field is 
\begin{eqnarray}
    F_{\tau r}=-\frac{i}{8\pi} \sqrt{(D-2)(D-3)}\frac{q}{r^{D-2}}\;, 
\end{eqnarray}
where there is an imaginary unit due to the time transformation $\tau=i t$. Note that there is also an imaginary unit $i$ in electromagnetic potential. The unit normal vector is $n_r=-1/\sqrt{f}$. After some algebra, one can finally get
\begin{eqnarray}
    I_{EM}=\frac{1}{16\pi} (D-2)\beta \Omega_{D-2} \frac{q^2}{r_h^{D-3}}\;.
\end{eqnarray}
It can also be written as 
\begin{eqnarray}
I_{EM}=\frac{\beta}{2}Q\Phi_H\;.
\end{eqnarray}

\subsection{Generalized free energy in canonical ensemble}

At last, considering all the contributions discussed above, we have the gravitational action as follows 
\begin{eqnarray}
I_{E}&=&\frac{1}{16\pi} (D-2)\beta \Omega_{D-2} \left(r_h^{D-3}+\frac{16\pi}{(D-1)(D-2)}P r_h^{D-1}+\alpha r_h^{D-5}+\frac{q^2}{r_h^{D-3}}\right)\nonumber\\
&&-\frac{1}{4} \Omega_{D-2} r_h^{D-2} \left(1+\frac{2\alpha(D-2)}{(D-4)}\frac{1}{r_h^2}\right)\;,
\end{eqnarray}
where we have expressed the mass parameter $\omega$ as the function of black hole radius $r_h$ and restored the cosmological constant as the thermodynamic pressure towards the relation $P=\frac{(D-1)(D-2)}{16\pi L^2}$.   

For the canonical ensemble, the free energy is defined in the semiclassical approximation as 
\begin{eqnarray}\label{F_SAdS}
F&=&-\frac{1}{\beta} \ln Z_{grav}(\beta) \nonumber\\
&=&\frac{I_E}{\beta} \nonumber\\
&=&
\frac{1}{16\pi} (D-2) \Omega_{D-2} \left(r_h^{D-3}+\frac{16\pi}{(D-1)(D-2)}P r_h^{D-1}+\alpha r_h^{D-5}+\frac{q^2}{r_h^{D-3}}\right)\nonumber\\
&&-\frac{1}{4} T \Omega_{D-2} r_h^{D-2} \left(1+\frac{2\alpha(D-2)}{(D-4)}\frac{1}{r_h^2}\right)\;.
\end{eqnarray}
This is just the generalized free energy of the fluctuating Gauss-Bonnet AdS black hole previously defined by using the thermodynamic relation $F=M-TS$ \cite{Wei:2020rcd,Wei:2021bwy}. To make it more explicit, we recall that the energy $E$ and entropy $S$ of a canonical ensemble at the temperature $T=1/\beta$ can be derived from the free energy as 
\begin{eqnarray}\label{E_S}
E&=&\frac{\partial}{\partial \beta}(\beta F)=\frac{1}{16\pi} (D-2) \Omega_{D-2} \left(r_h^{D-3}+\frac{16\pi}{(D-1)(D-2)}P r_h^{D-1}+\alpha r_h^{D-5}+\frac{q^2}{r_h^{D-3}}\right)\;,\\
S&=&\beta(E-F)=\frac{1}{4}  \Omega_{D-2} r_h^{D-2} \left(1+\frac{2\alpha(D-2)}{(D-4)}\frac{1}{r_h^2}\right)\;.
\end{eqnarray}
It can be seen that the energy $E$ and the entropy $S$ are independent of the inverse temperature $\beta$. By identifying the energy $E$ as the black hole mass $M$, the thermodynamic definition of the generalized free energy is then given by 
\begin{eqnarray}\label{F_MTS}
F=M-TS&=&\frac{1}{16\pi} (D-2) \Omega_{D-2} \left(r_h^{D-3}+\frac{16\pi}{(D-1)(D-2)}P r_h^{D-1}+\alpha r_h^{D-5}+\frac{q^2}{r_h^{D-3}}\right)\nonumber\\
&&-\frac{1}{4} T \Omega_{D-2} r_h^{D-2} \left(1+\frac{2\alpha(D-2)}{(D-4)}\frac{1}{r_h^2}\right)
\end{eqnarray}
which coincides with the result Eq.(\ref{F_SAdS}) calculated from the gravitational action on the singular Euclidean manifold as instanton. In this expression, the generalized free energy should be considered as the function of the black hole radius $r_h$.

\section{free energy landscape in diverse spacetime dimensions}\label{SecIV}

In this section, we will discuss the free energy landscape of the Gauss-Bonnet black holes in diverse spacetime dimensions. Free energy landscape is an intuitive representation of the generalized free energy function. Free energy landscape can have different shapes at different ensemble temperatures. In general, for the first order phase transition of the small/large Gauss-Bonnet black holes, the free energy landscape has of the shape of double well.   

\subsection{critical point from generalized free energy}

Firstly, let us discuss how to determine the critical point from the generalized free energy function. Hawking temperature has special meaning on the free energy landscape. When the ensemble temperature is equal to the Hawking temperature, the black hole is in equilibrium state with the thermal bath. In this case, the black is locally stable. The thermodynamically stable black hole states correspond to the local extreme points of the generalized free energy function. They are determined by the equation
\begin{eqnarray}\label{pGibbs}
\frac{\partial F}{\partial r_h}=\frac{(D-2)}{16\pi} \Omega_{D-2} \left((D-3)r_h^{D-4}+ \frac{16\pi P r_h^{D-2}}{(D-2)} +(D-5)\alpha r_h^{D-6}\right.\nonumber\\ \left.-(D-3) \frac{q^2}{r_h^{D-2}} -4\pi T r_h^{D-5} (r_h^2+2\alpha)\right)=0\;.
\end{eqnarray}
This equation gives us the relation between the ensemble temperature $T$ and the black hole radius $r_h$ when the black hole is in equilibrium. It is easy to check that solving the ensemble temperature $T$ from the above equation gives the expression of the Hawking temperature $T_H$ as shown in Eq.(\ref{Hawk_temp}).

In general, it can be observed that when the ensemble temperature is in a specific range, the free energy landscape has the shape of a double well. This is to say that there exist a minimal temperature $T_{min}$ and a maximum temperatures $T_{max}$. When $T<T_{min}$ and $T>T_{max}$, the shape of the free energy landscape is a single well and no phase transition can occur. When $T_{min}<T<T_{max}$, the free energy landscape is double well and there is a first order phase transition between the small and the large black holes. If the minimal and the maximum temperatures coincide, i.e. $T_{min}=T_{max}$, the system lies at the critical point.

On the free energy landscape, the appearance or the disappearance of the double well shape corresponds to the equation $\frac{\partial^2 F}{\partial r_h^2}=0$, which also indicates that there is an inflection point. The minimal temperature $T_{min}$ and the maximum temperature $T_{max}$ are then determined by the following equation
\begin{eqnarray}\label{ppGibbs}
\frac{\partial^2 F}{\partial r_h^2}=\frac{(D-2)}{16\pi} \Omega_{D-2}\left( (D-3)(D-4)r_h^{D-5} +16 \pi  P r_h^{D-3}
+ (D-5) (D-6) \alpha r_h^{D-7} \right.\nonumber\\ 
\left.
+(D-2) (D-3)\frac{q^2}{r_h^{D-1}} -4 \pi T \left((D-3) r_h^{D-4}+2 \alpha  (D-5) r_h^{D-6}\right) \right)=0\;.
\end{eqnarray}
Combining Eq.(\ref{pGibbs}) and Eq.(\ref{ppGibbs}) and eliminating $T$, we can get the equation
\begin{eqnarray}\label{crit_eq}
\frac{16\pi P}{(D-2)} r_h^3 \left(r_h^2+6\alpha \right)-(D-3) r_h^3- (D-9)\alpha r_h + 2(D-5)\frac{\alpha ^2}{r_h}\nonumber\\ +(D-3)\frac{ q^2}{r_h^{2D-7}} \left((2 D-5) r_h^2+2(2D-7) \alpha \right)=0\;.
\end{eqnarray}
When the minimal and the maximum temperatures coincide at the critical point, the above equation has two equal roots. This condition gives us the critical pressure, the critical temperature, as well we the critical black hole radius. However, this equation is highly nonlinear but the numerical solution can be easily obtained.  

\subsection{D=5}

For $D=5$, the generalized free energy can be explicitly given by
\begin{eqnarray}\label{F_d5}
F=\frac{3\pi}{8}\left(r_h^2+\frac{4\pi}{3}P r_h^{4}+\alpha +\frac{q^2}{r_h^2}\right)-\frac{\pi^2}{2} T r_h^{3} \left(1+\frac{6\alpha}{r_h^2}\right)\;.
\end{eqnarray}
In analogy to the liquid-gas phase transition, there is also a critical point on the phase diagram for the small/large Gauss-Bonnet black hole state switching and phase transition. As previously discussed, the critical point can be determined numerically. For $q=1$ and $\alpha=0.1$, the critical pressure is $P_c=0.0165$, the critical temperature is $T_c=0.113$, and the critical black hole radius is $r_h^c=2.084$. When the pressure is below the critical pressure, there is a small/large black hole state switching and phase transition in the Gauss-Bonnet gravity system.  

\begin{figure}
  \centering
  \includegraphics[width=8cm]{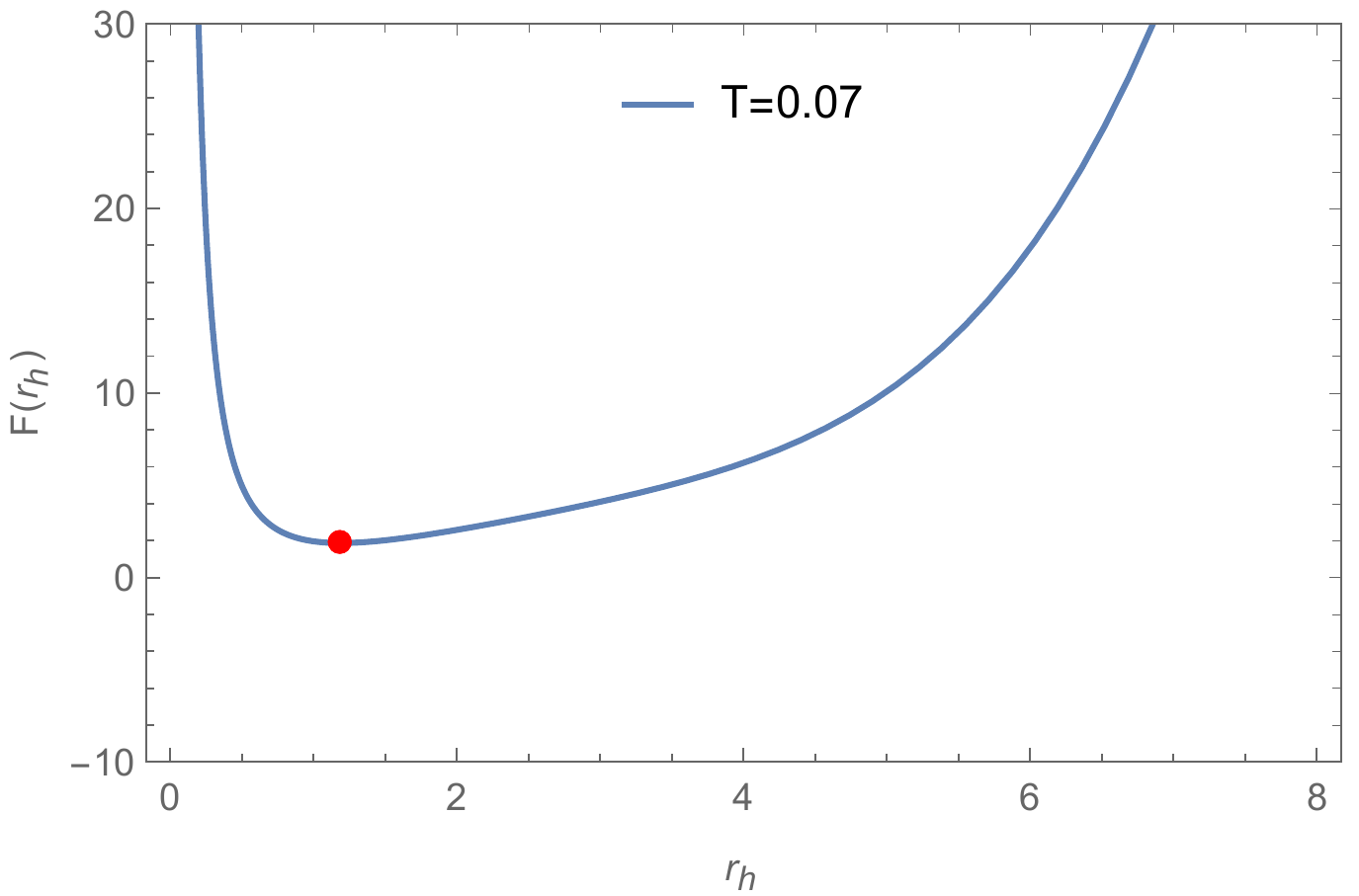}
  \includegraphics[width=8cm]{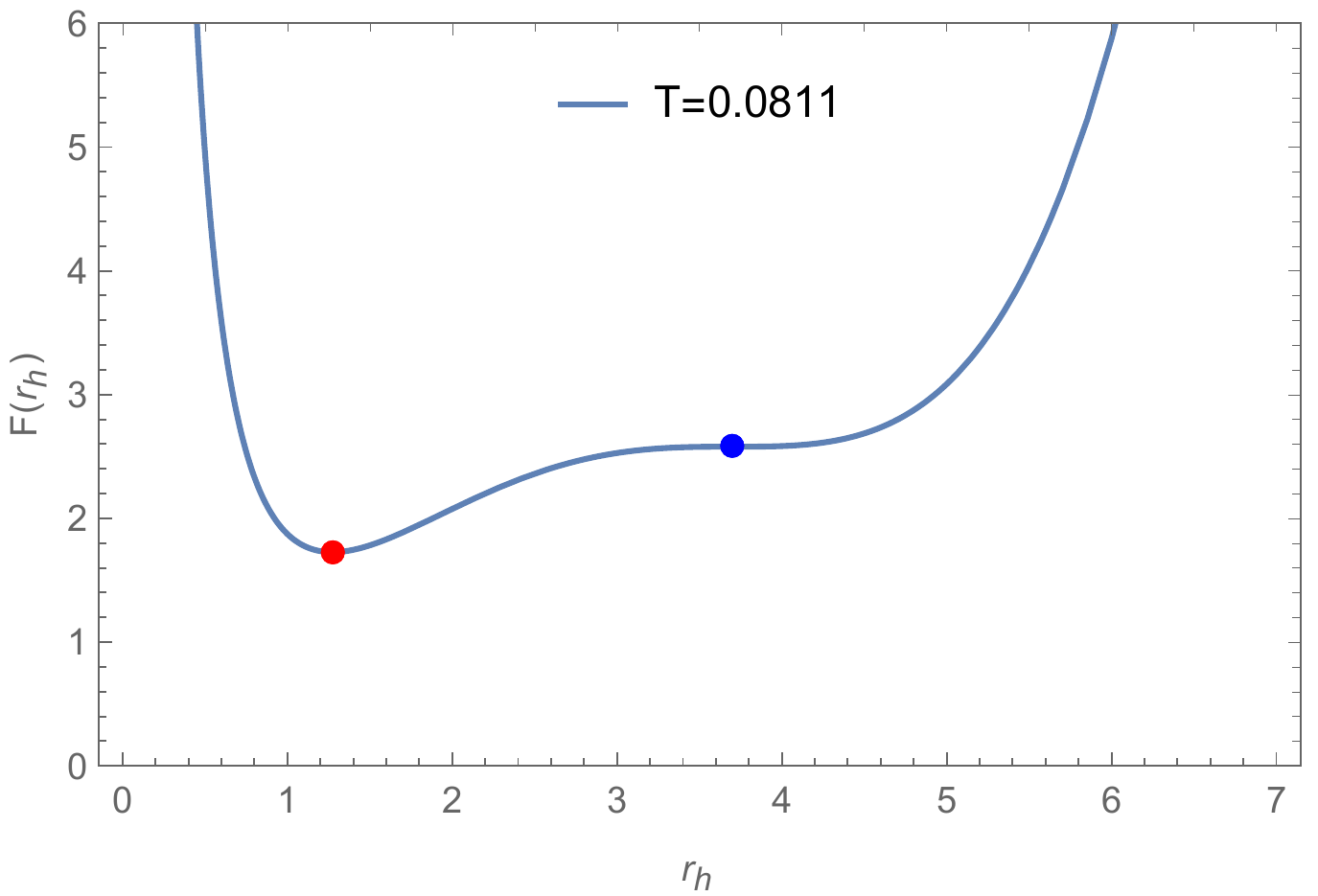}\\
  \includegraphics[width=8cm]{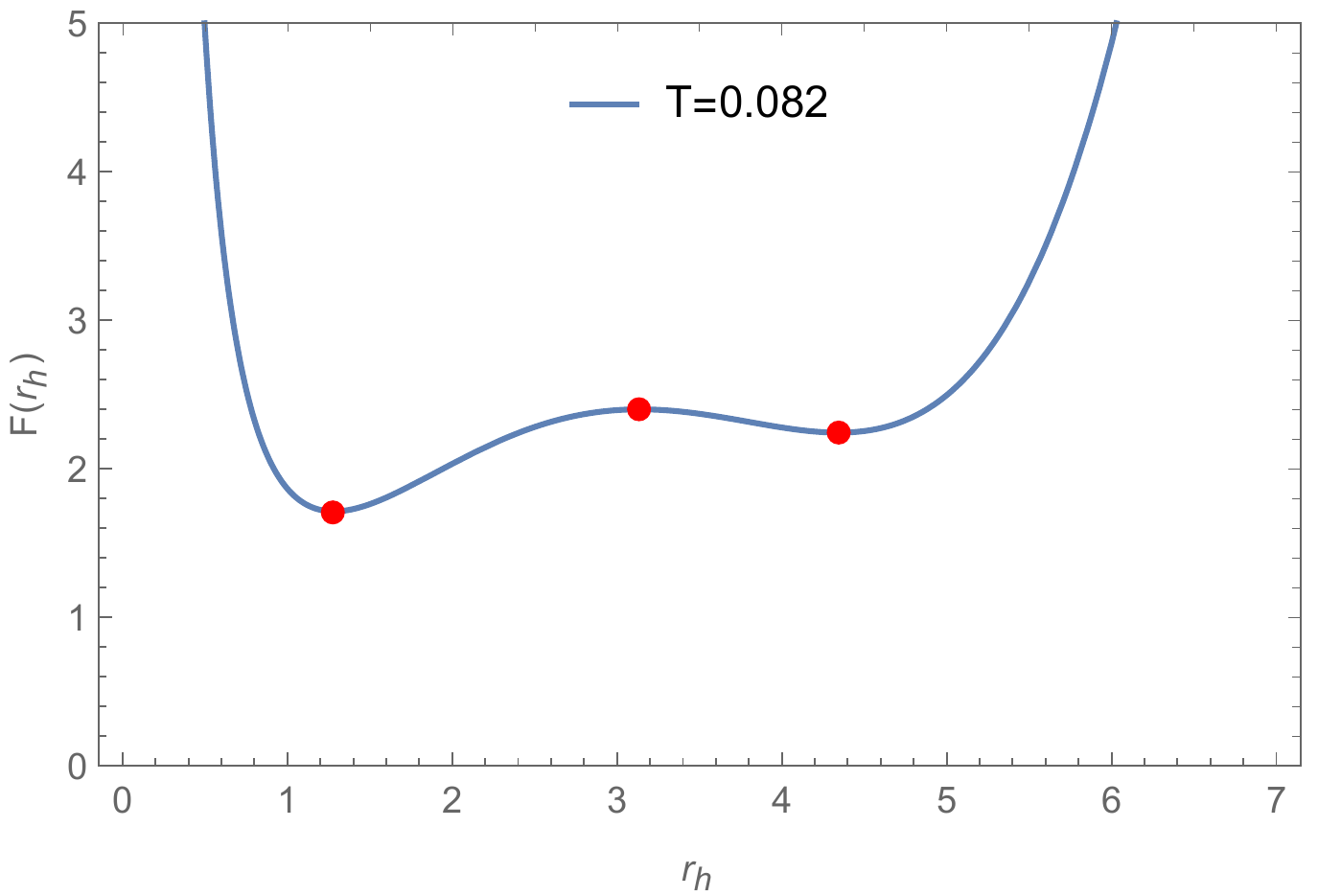}
  \includegraphics[width=8cm]{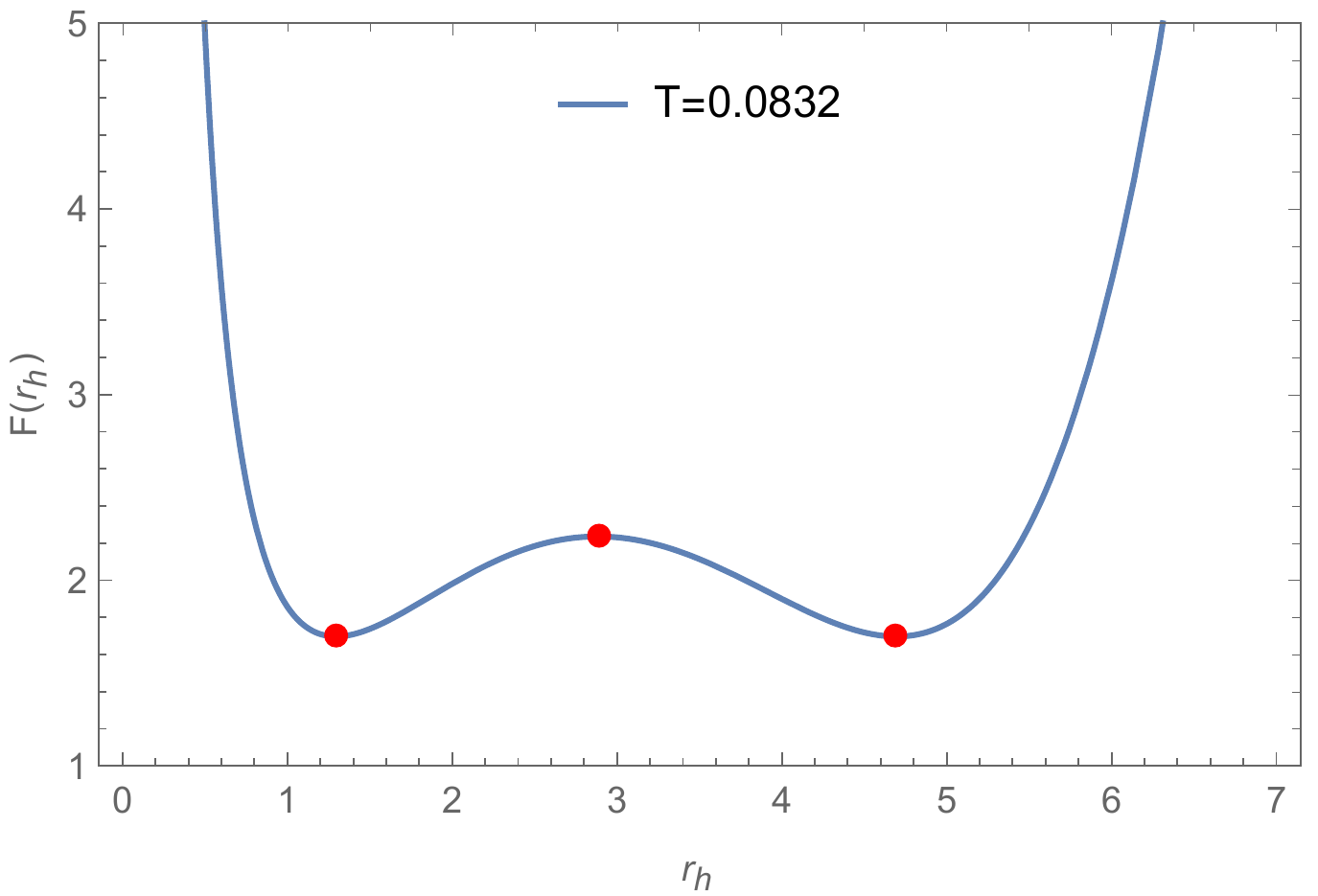}\\
  \includegraphics[width=8cm]{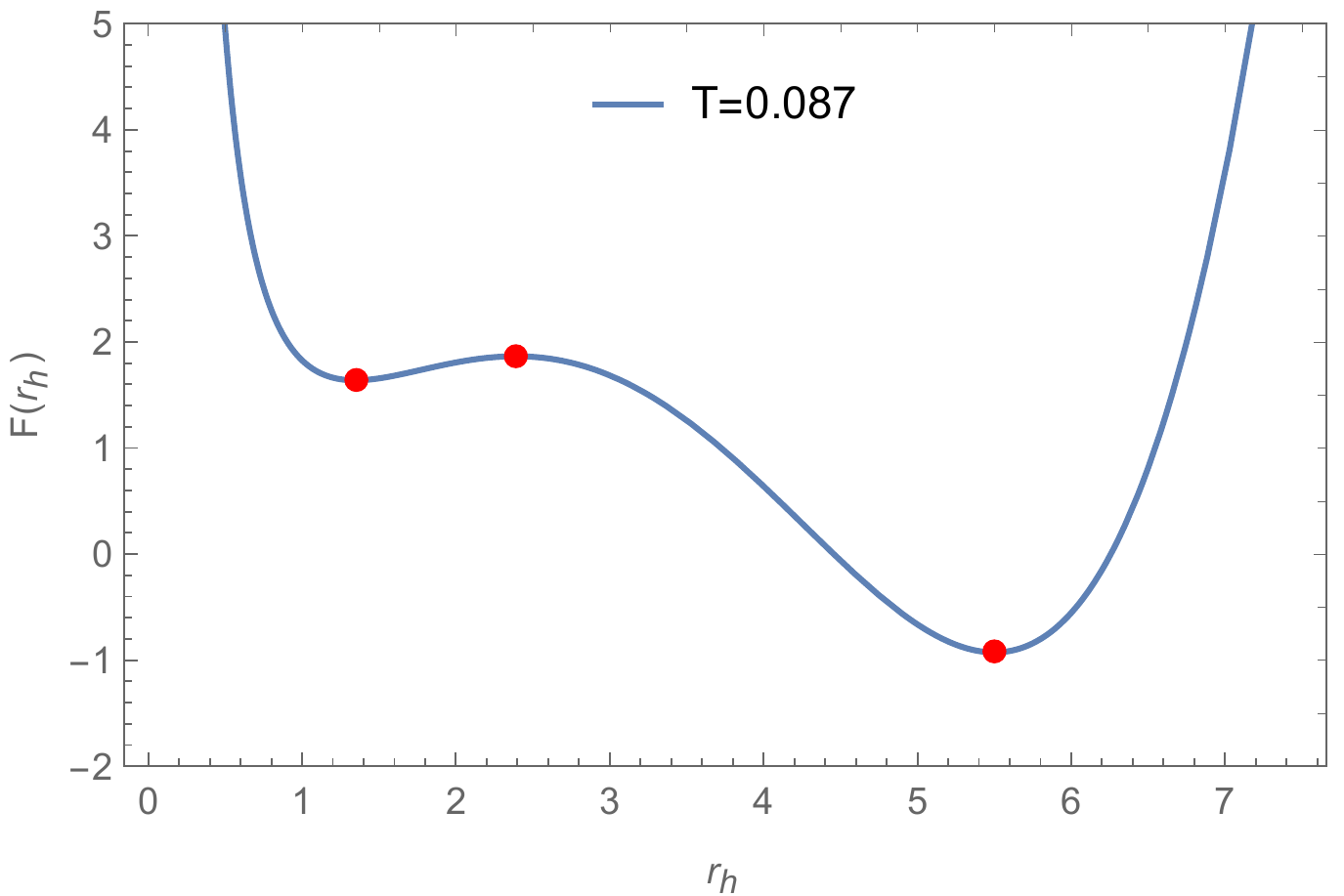}
  \includegraphics[width=8cm]{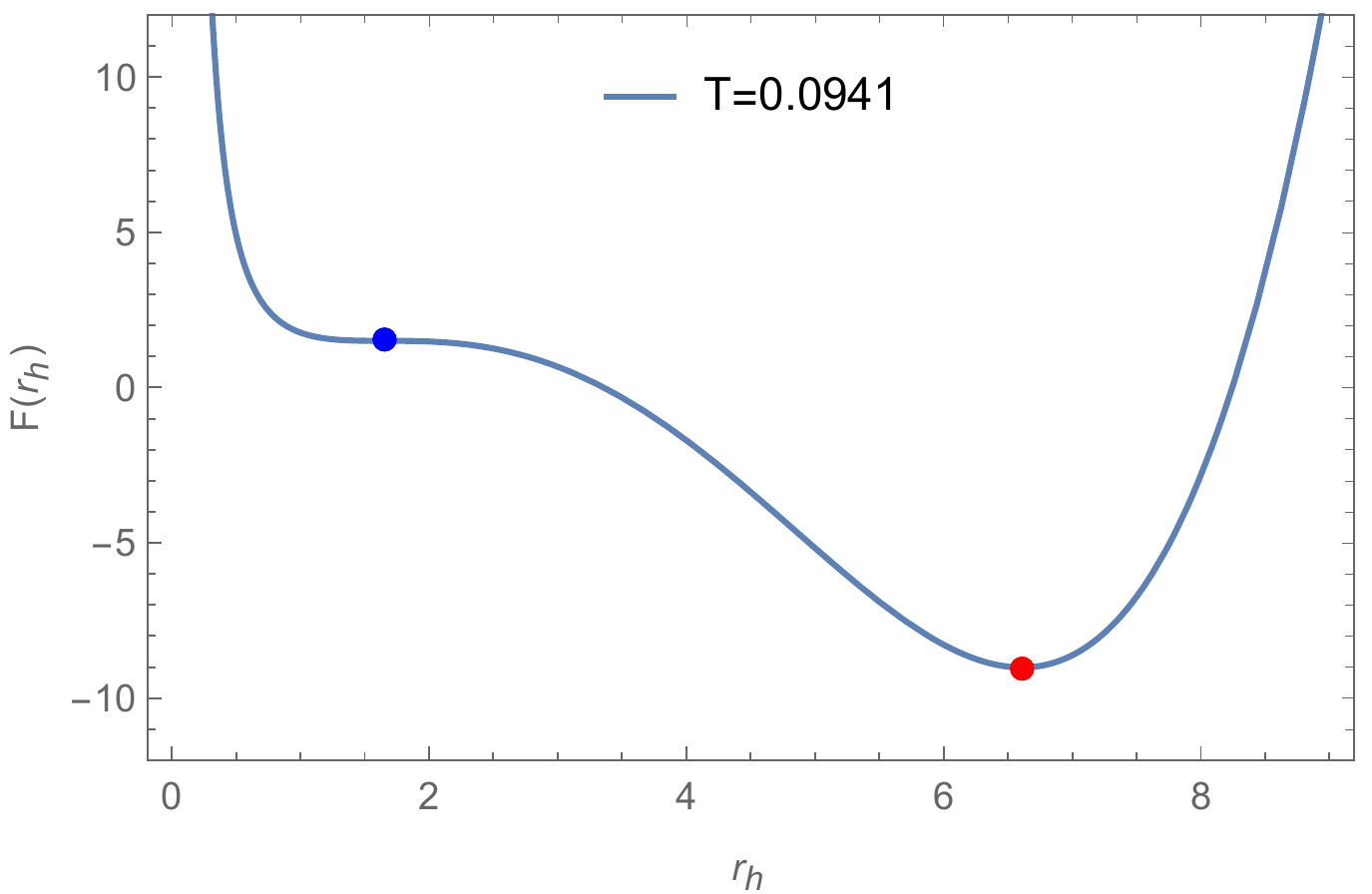}\\
  \caption{Free energy landscape for $D=5$. In this plot, $P=0.008$, $\alpha=0.1$, and $q=1$. When $0.08686<T<0.1472$, the landscape is of the shape of double well. Otherwise, it is single well. The red points represent the Gauss-Bonnet AdS black holes in the equilibrium state, while the blue points represent the inflection points on the landscape. }
  \label{Landscape_d5}
\end{figure}

Free energy landscape is a powerful tool for studying the thermodynamics of phase transition. In Fig.\ref{Landscape_d5}, we have plotted the generalized free energy as the function of black hole radius for $P=0.008$. At different temperatures, the landscapes takes on different shapes, as illustrated.

At $T=0.07$, there is only one stationary state (represented by the red point) on the landscape, which represents the stable Gauss-Bonnet black hole. In this case, there is no phase transition. As the temperature increases to $T=0.0811$, an inflection point (blue) appears on the free energy landscape along with the stationary point. The stationary point still represents the stable black hole state, while the inflection point marks the emergence of the large black hole and the intermediate black hole states in the Gauss-Bonnet gravity system. When the temperature further increases to $T=0.082$, the landscape takes on a double well shape with two locally stable states and one unstable state, all represented by the red points on the landscape. In this case, the free energy of the left stationary point is lower than that for the right stationary point, indicating that the small black hole state represented by the left stationary point is the globally stable state in thermodynamics. At the phase transition point $T=0.0832$, the potential depths of the left and right stationary points become equal, allowing the small and large black hole states to coexist. As the temperature continues to rise to $T=0.087$, the similar analysis can be performed to obtain that the large black hole state is the globally stable one. Finally, at $T=0.0941$, another inflection point appears, and the shape of the landscape begins to restore the shape of a single well. This describes the dependence of the thermodynamics of the black hole state switching and phase transition on the ensemble temperature.

\begin{figure}
  \centering
  \includegraphics[width=8cm]{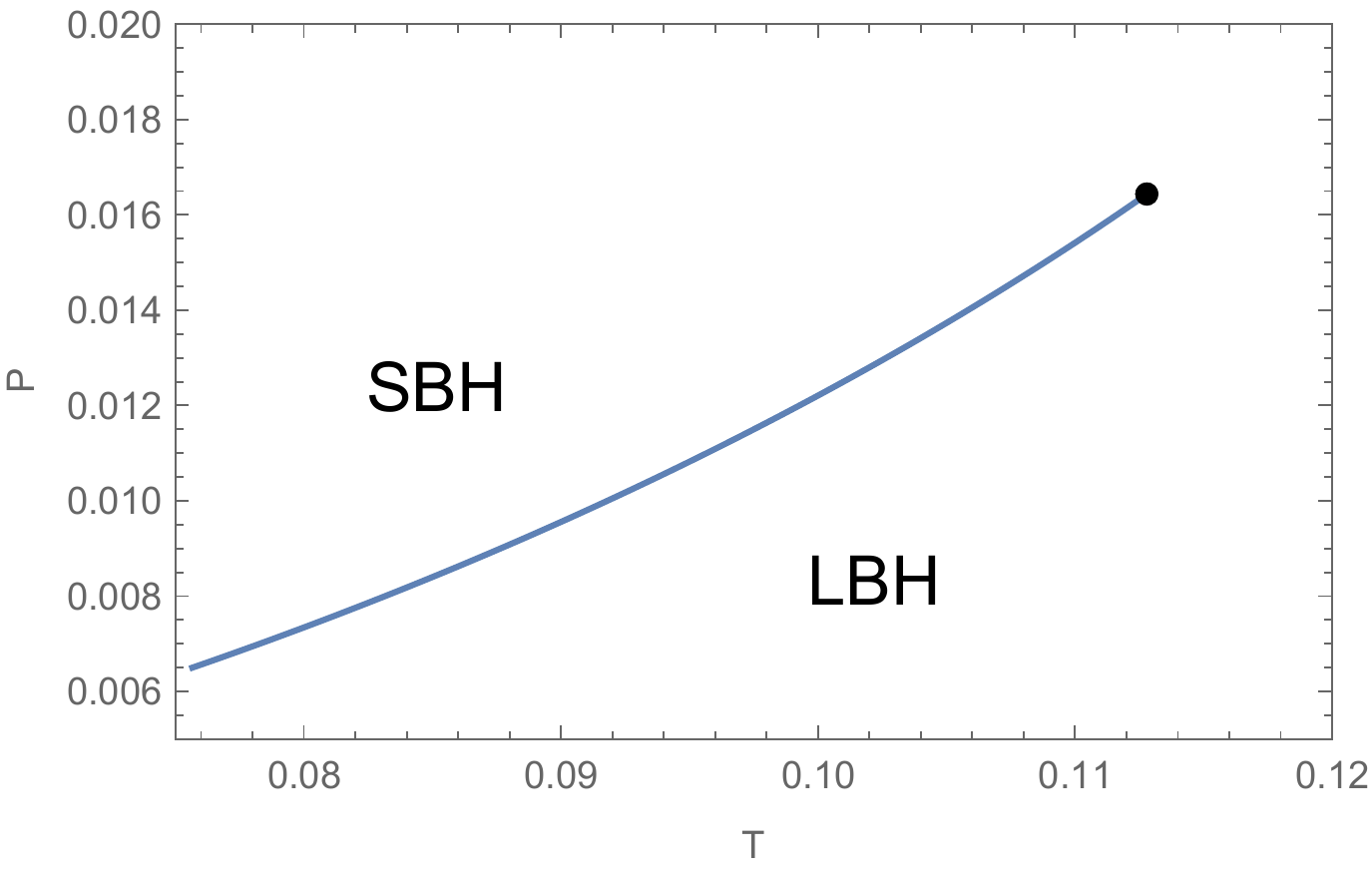}\\
  \caption{Phase diagram for $D=5$. In this plot, $\alpha=0.1$, and $q=1$. The black point represents the critical point. "SBH" and "LBH" represent the small and the large Gauss-Bonnet black holes respectively. The curve is the coexisting curve of the small and the large black holes.} 
  \label{Phase_diagram_d5}
\end{figure}

Based on the discussion of the influence of the ensemble temperature on the free energy landscapes, we have also studied the phase structure of the charged Gauss-Bonnet AdS black holes in five dimensions by plotting the coexisting curve on the ``T-P" plane (refer to Fig.\ref{Phase_diagram_d5}). For a fixed thermodynamic pressure that is below the critical value, one can numerically determine the ensemble temperature at which the two wells on the free energy landscape have the same depths. By varying the pressure, one can obtain the coexisting curve on the ``T-P" plane numerically. Therefore, the coexisting curve marks the phase transition point at which both the small and large black holes can exist simultaneously. This curve terminates at the critical point, which is denoted by black point on the phase diagram. It is also marked that for the region above/below the coexisting curve, the small/large Gauss-Bonnet AdS black hole is the globally stable state. As a result of our analysis, we have gained insight into the behavior of the charged Gauss-Bonnet black holes and their thermodynamic properties from the generalized free energy and the corresponding landscapes. 

\subsection{D=6}

For $D=6$, the generalized free energy is given by 
\begin{eqnarray}
F=\frac{2\pi}{3}   \left( r_h^3+ \frac{4\pi}{5}  P r_h^5+\alpha r_h+\frac{q^2}{r_h^3}\right)-\frac{2\pi^2}{3} T  r_h^4  \left(1+ \frac{4 \alpha }{r_h^2}\right)\;.
\end{eqnarray}

The critical point can also be easily obtained by solving Eq.(\ref{crit_eq}) numerically. In the following, we will discuss two cases: one critical point and two critical points.

\begin{figure}
  \centering
  \includegraphics[width=8cm]{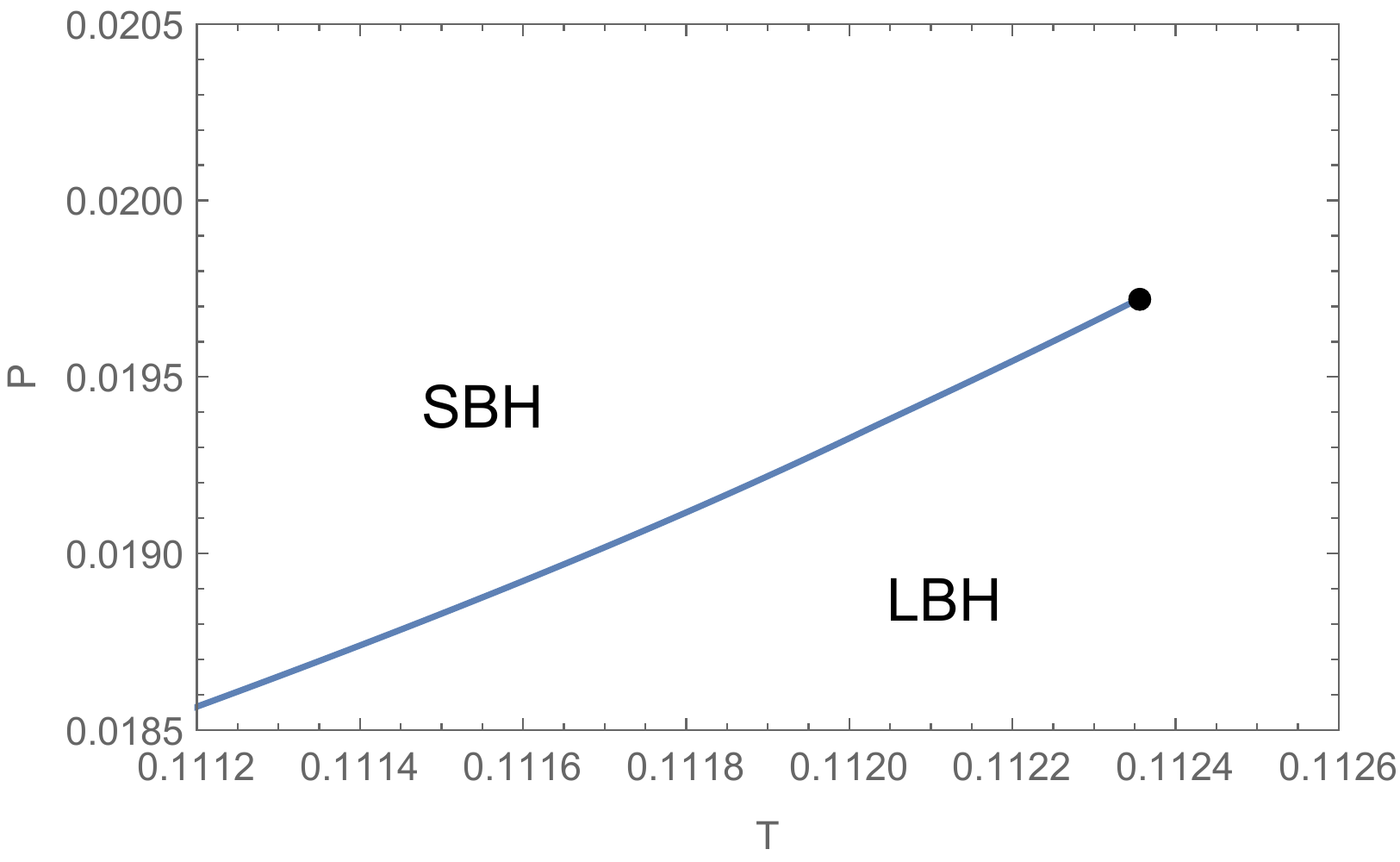}\\
  \caption{Phase diagram for $D=6$, with $q=0.08$ and $\alpha=1$. The black point represents the critical point and the curve is the coexisting curve. "SBH" and "LBH" represent the small and the large Gauss-Bonnet black holes respectively. }
  \label{Phase_diagram_d6_trivial}
\end{figure}

\begin{figure}
  \centering
  \includegraphics[width=8cm]{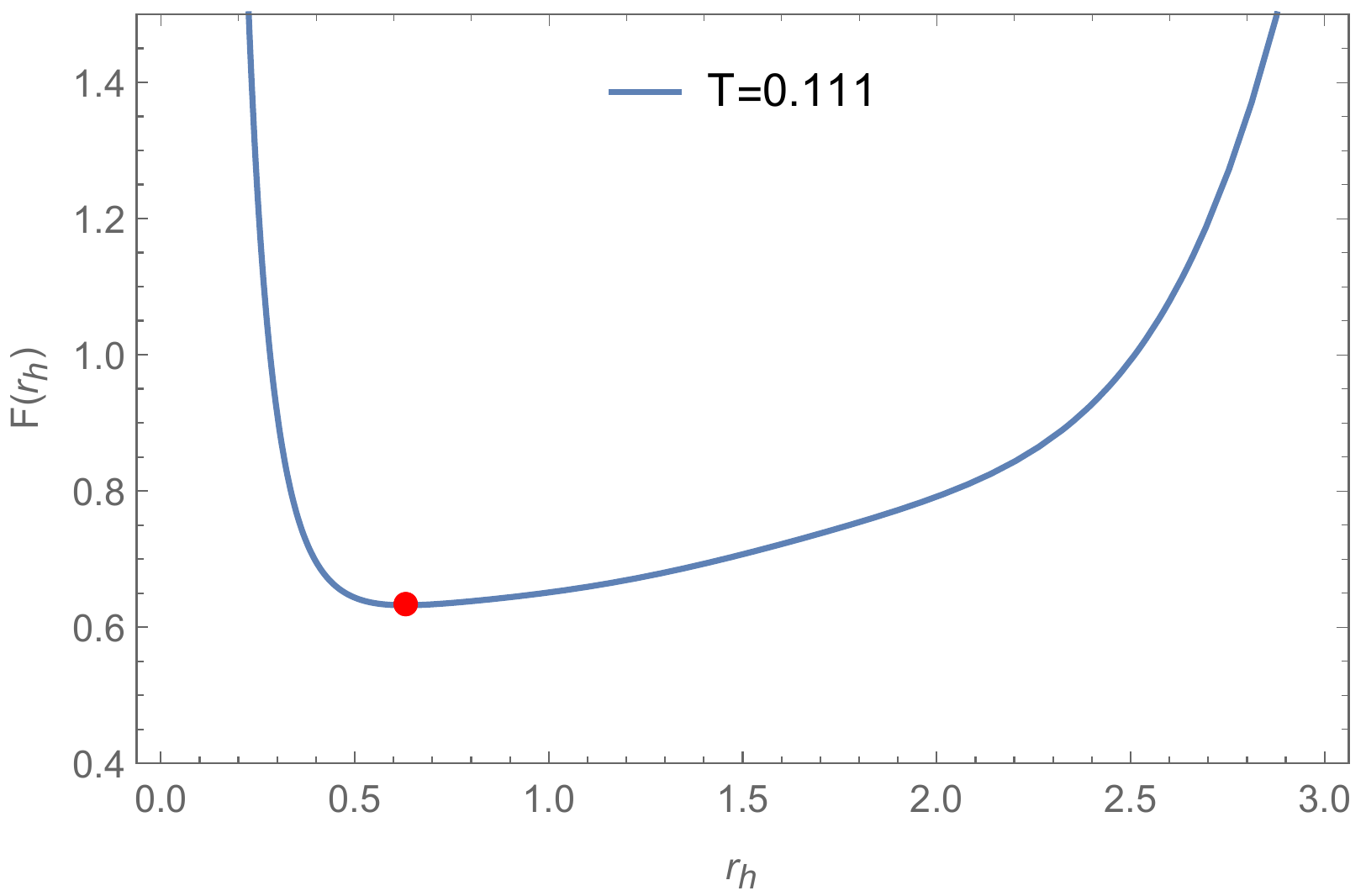}
  \includegraphics[width=8cm]{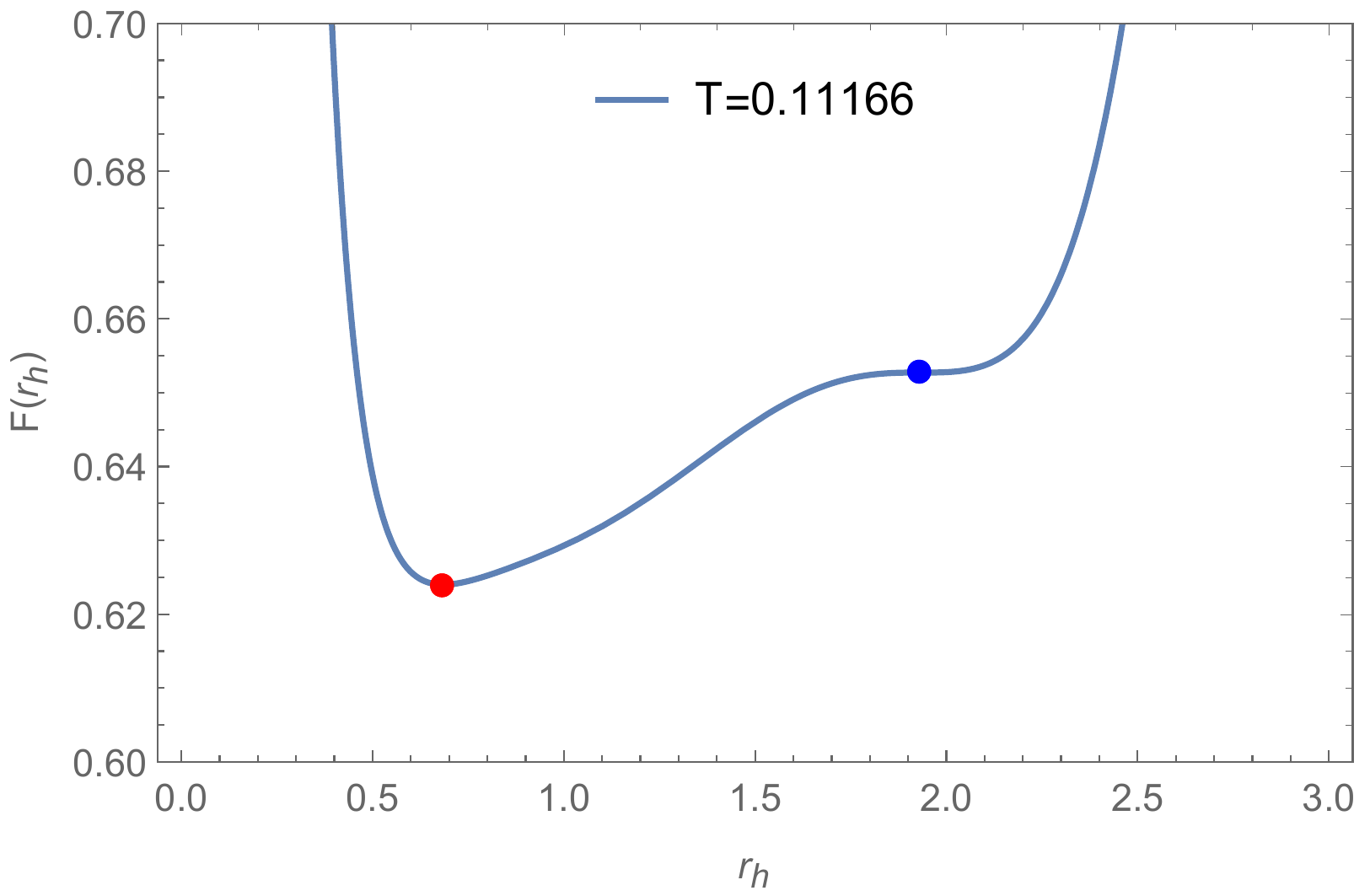}\\
  \includegraphics[width=8cm]{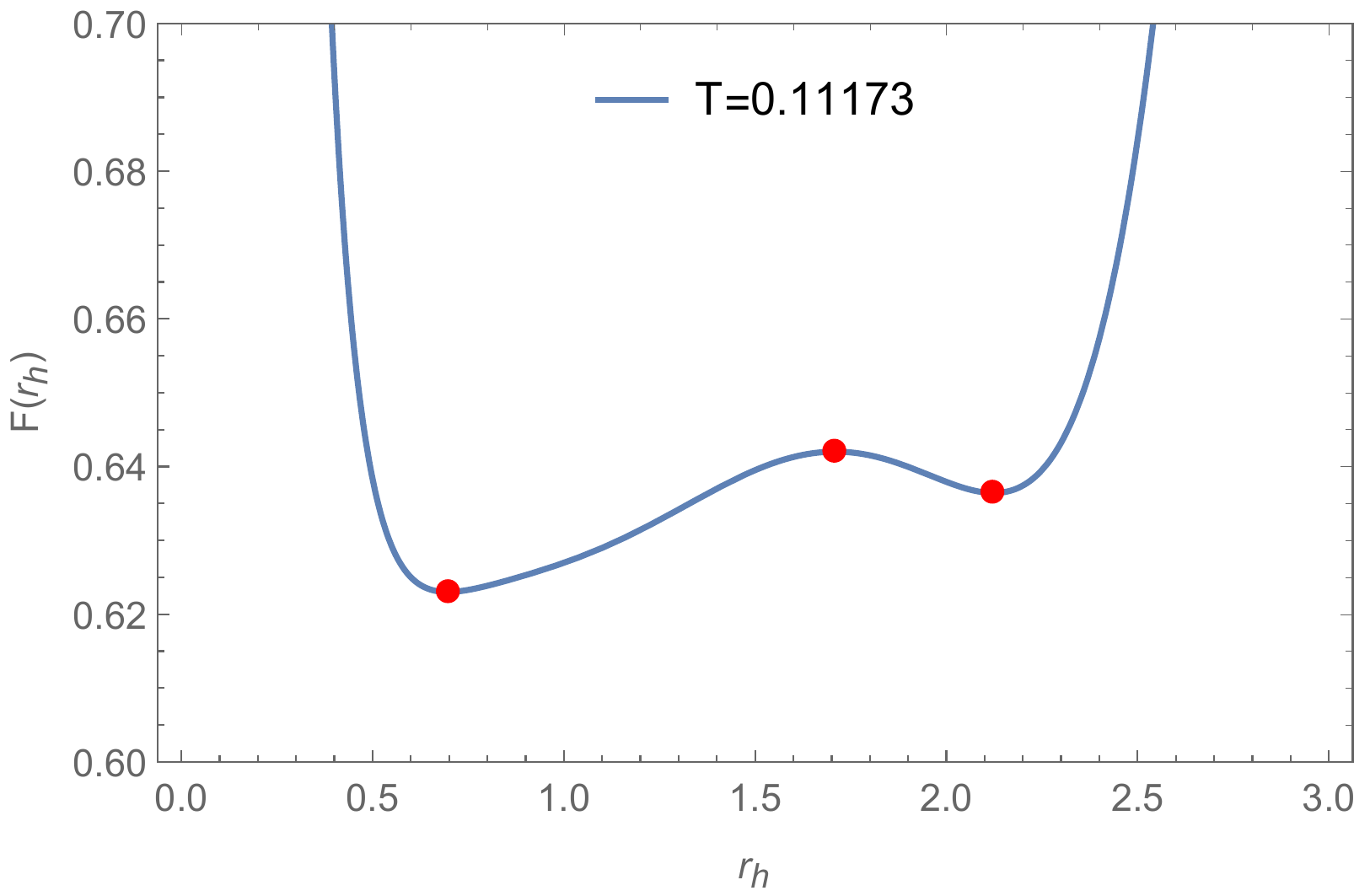}
  \includegraphics[width=8cm]{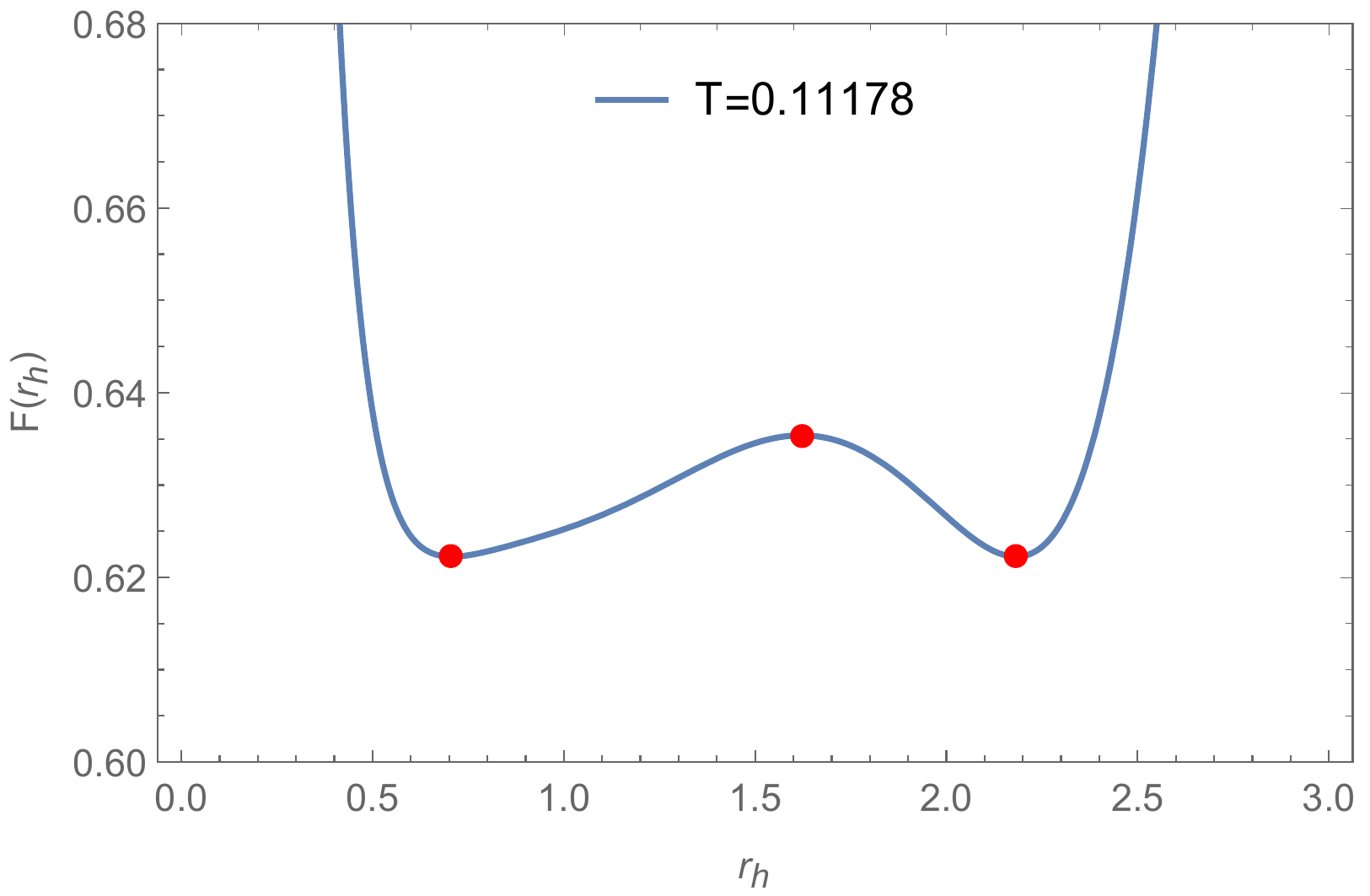}\\
  \includegraphics[width=8cm]{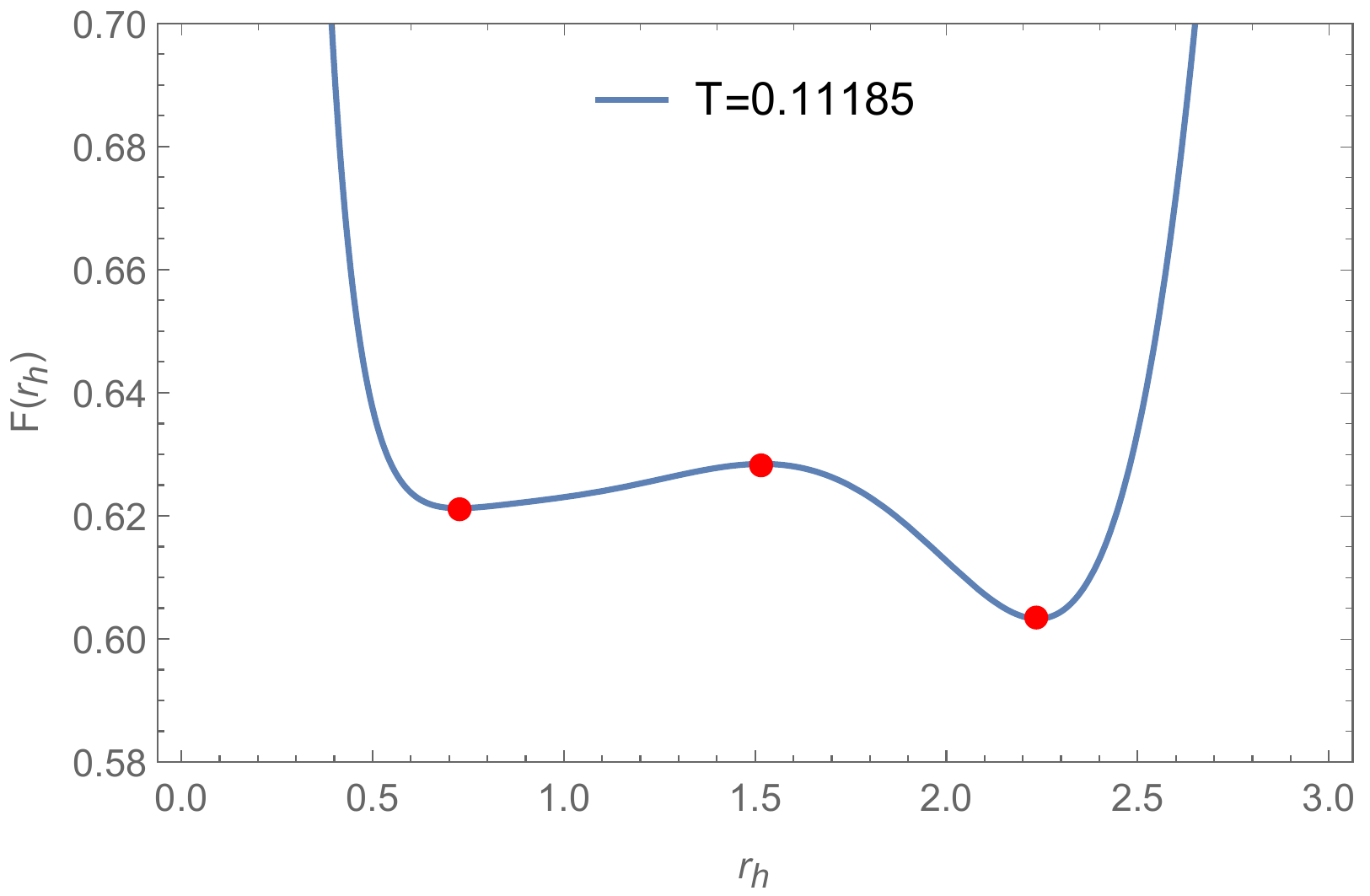}
  \includegraphics[width=8cm]{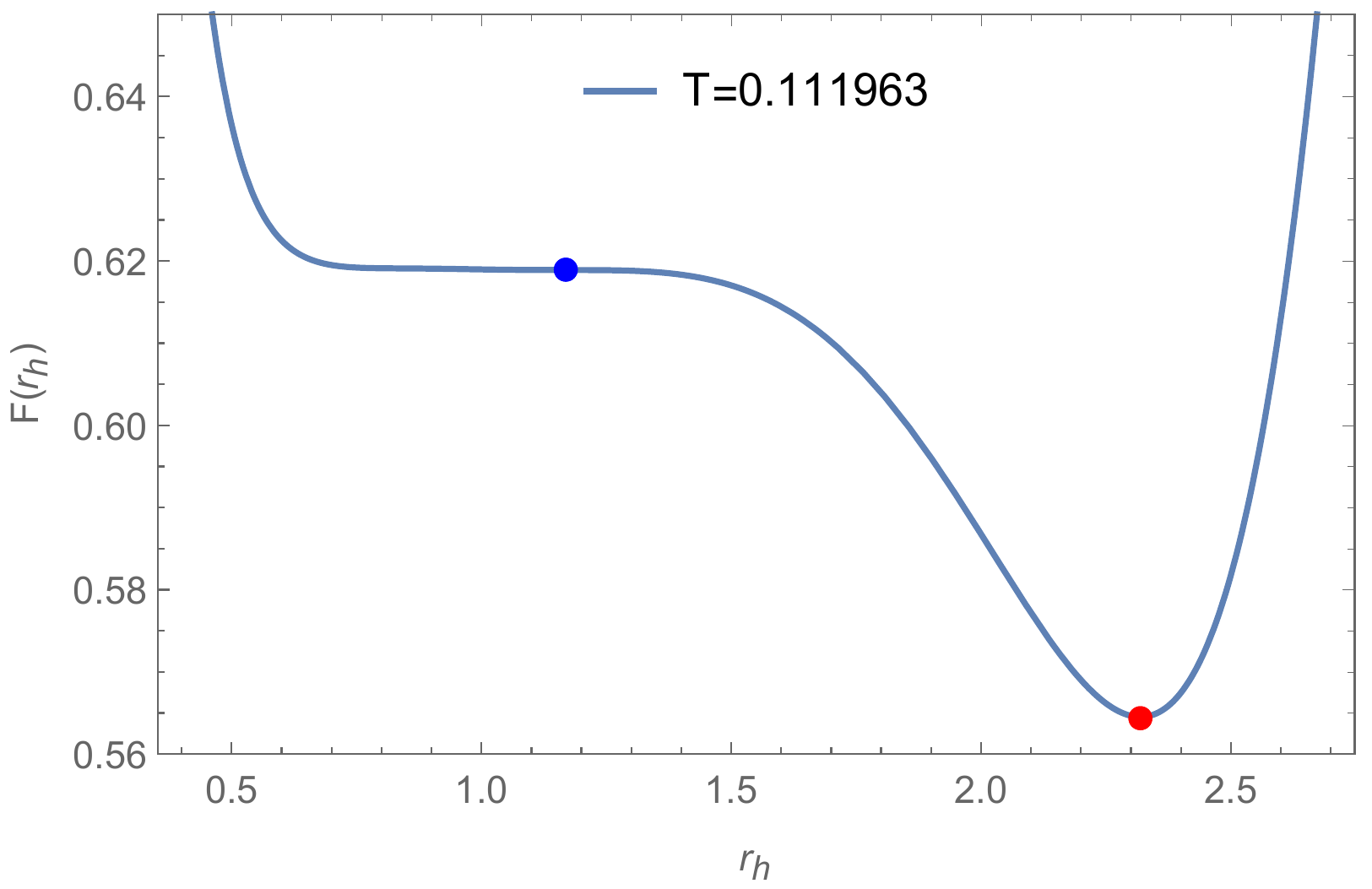}\\
  \caption{Free energy landscape for $D=6$, with $P=0.008$, $\alpha=1$ and $q=0.08$. The landscapes are of the shapes of double well when $0.11166<T<0.111963$. Otherwise, it is single well. The phase transition critical temperature is $T=0.11178$. The red points represent the  Gauss-Bonnet AdS black holes in the equilibrium state, while the blue points represent the inflection points on the landscape. }
  \label{Landscape2_d6}
\end{figure}

\begin{figure}
  \centering
  \includegraphics[width=8cm]{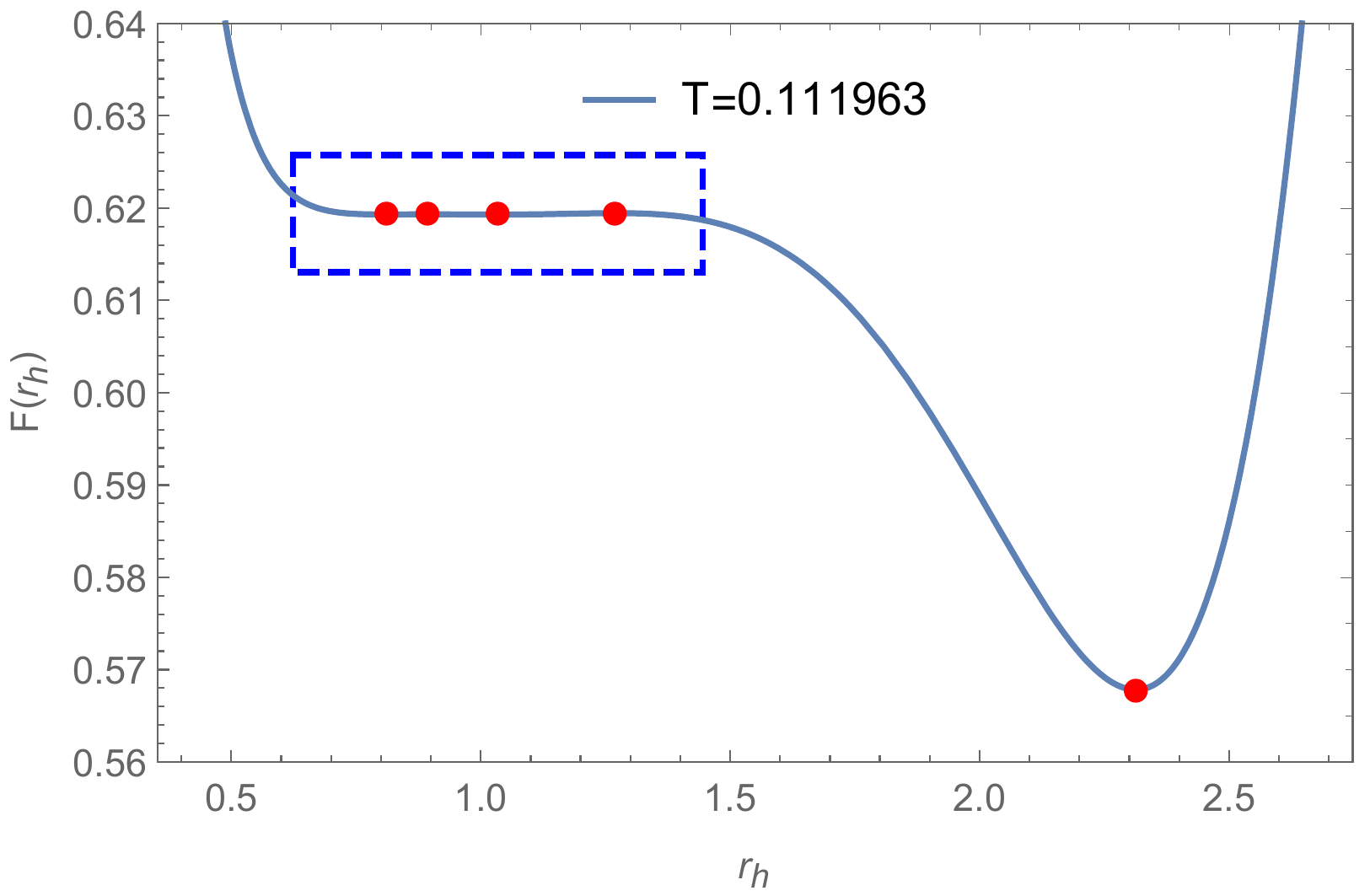}
  \includegraphics[width=8cm]{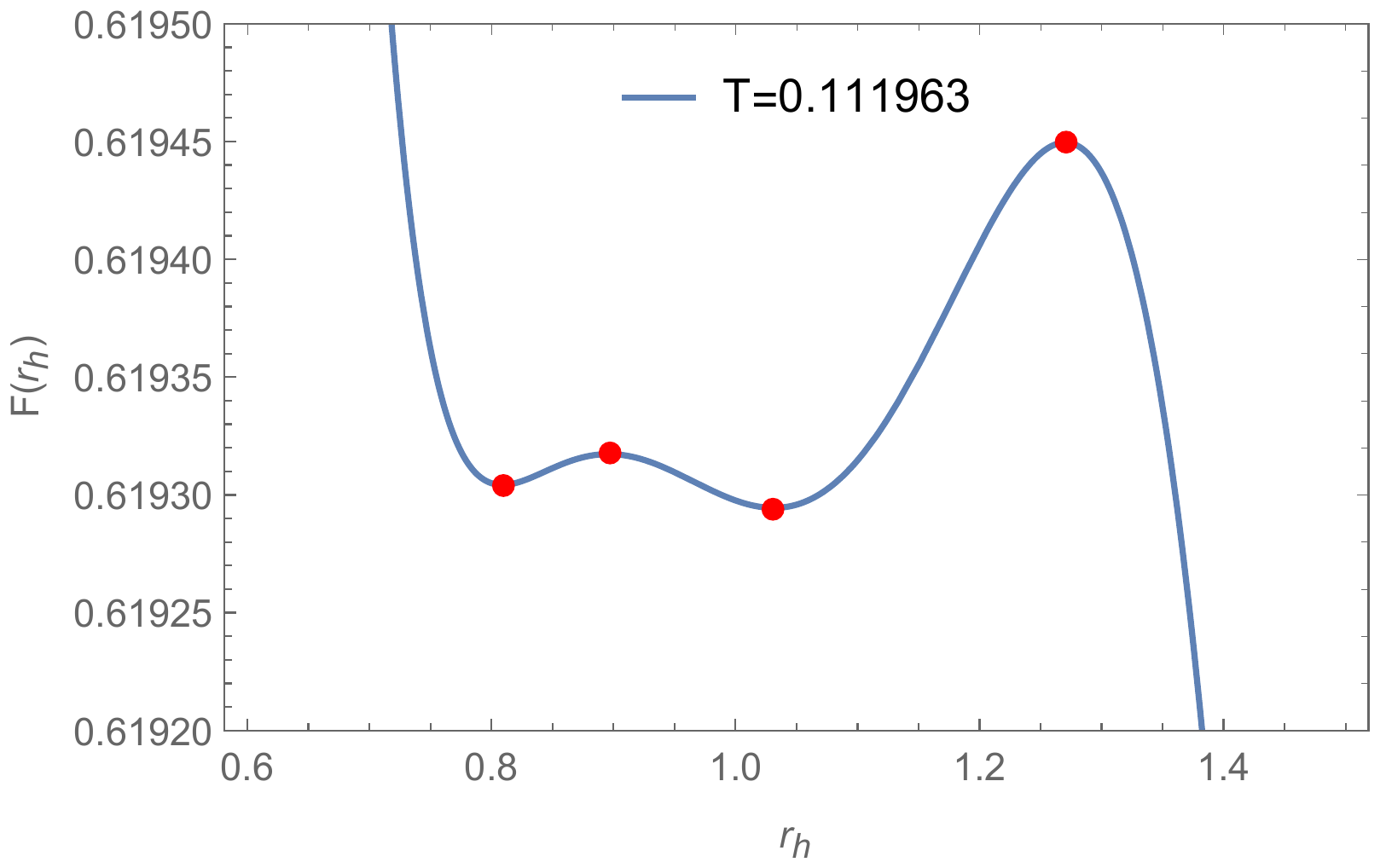}\\
  \caption{Free energy landscape for $D=6$. In this plot, $P=0.008$, $\alpha=1$, and $q=0.08$. The right panel is the enlarged plot of the rectangular region in the left panel. When $0.1119595<T<0.1119668$, the landscape is of the shape of three well.}
  \label{Landscape3_d6_ST}
\end{figure}

We firstly consider the there is only one critical point on the phase diagram, which is plotted in Fig.\ref{Phase_diagram_d6_trivial}. The phase diagram can be numerically obtained as discussed for $D=5$ case. It is shown that for $q=0.08$ and $\alpha=1$, the critical pressure is $P_c=0.01972$ and the critical temperature is $T_c=0.11236$. In this case, there are only two Gauss-Bonnet AdS black hole phases which will dominate when varying the thermodynamics parameters, although additional two black hole states will emerge at a very small temperature range. The dominated black hole state is always the small or the large Gauss-Bonnet AdS black hole. The additional two black hole states emerge at the ensemble temperature where the large Gauss-Bonnet AdS black hole is the thermodynamically stable state. To see this point more explicitly, we present the free energy landscapes for different ensemble temperatures in Fig.\ref{Landscape2_d6} and Fig.\ref{Landscape3_d6_ST}.

We show the shapes of the landscapes for $D=6$ at different ensemble temperatures in Fig.\ref{Landscape2_d6}. Compared with the free energy landscapes for $D=5$ case, the varying of the shapes of landscapes along with the ensemble temperature seems trivial. When the temperature is low, the shape of the landscape is a single well. As the temperature increases, the double well' shapes of the landscapes begin to emerge and then disappears at a specific temperature. Finally, it restores the single well's shape. However, this is not the full story.

A detailed analysis indicates that in a very narrow temperature range, i.e. when $0.1119595<T<0.1119668$, the landscapes have the shapes of three well, as depicted in Fig.\ref{Landscape3_d6_ST}. In the left panel, it can be observed that there are five locally extremal points (indicated by red points), which is different from the case of double well's landscapes shown in Fig.\ref{Landscape2_d6}. The right panel is an enlarged plot of the rectangular region in the left panel. It is cleat that there are three wells on the landscape. The large Gauss-Bonnet AdS black hole represented by the minimum point in the rightmost well is the thermodynamically stable state. Within the rectangular region of the left panel, there are four extremal points, indicating the emergence of additional two black hole phases. one of the emerged phases is unstable and the other phase is locally stable. However, the emerged locally stable phase is not the globally stable one since its free energy is always greater than that of the large black hole. In this case, there is no triple point where three phases coexist, and therefore, the two emerged phases will disappear quickly.

\begin{figure}
  \centering
  \includegraphics[width=8cm]{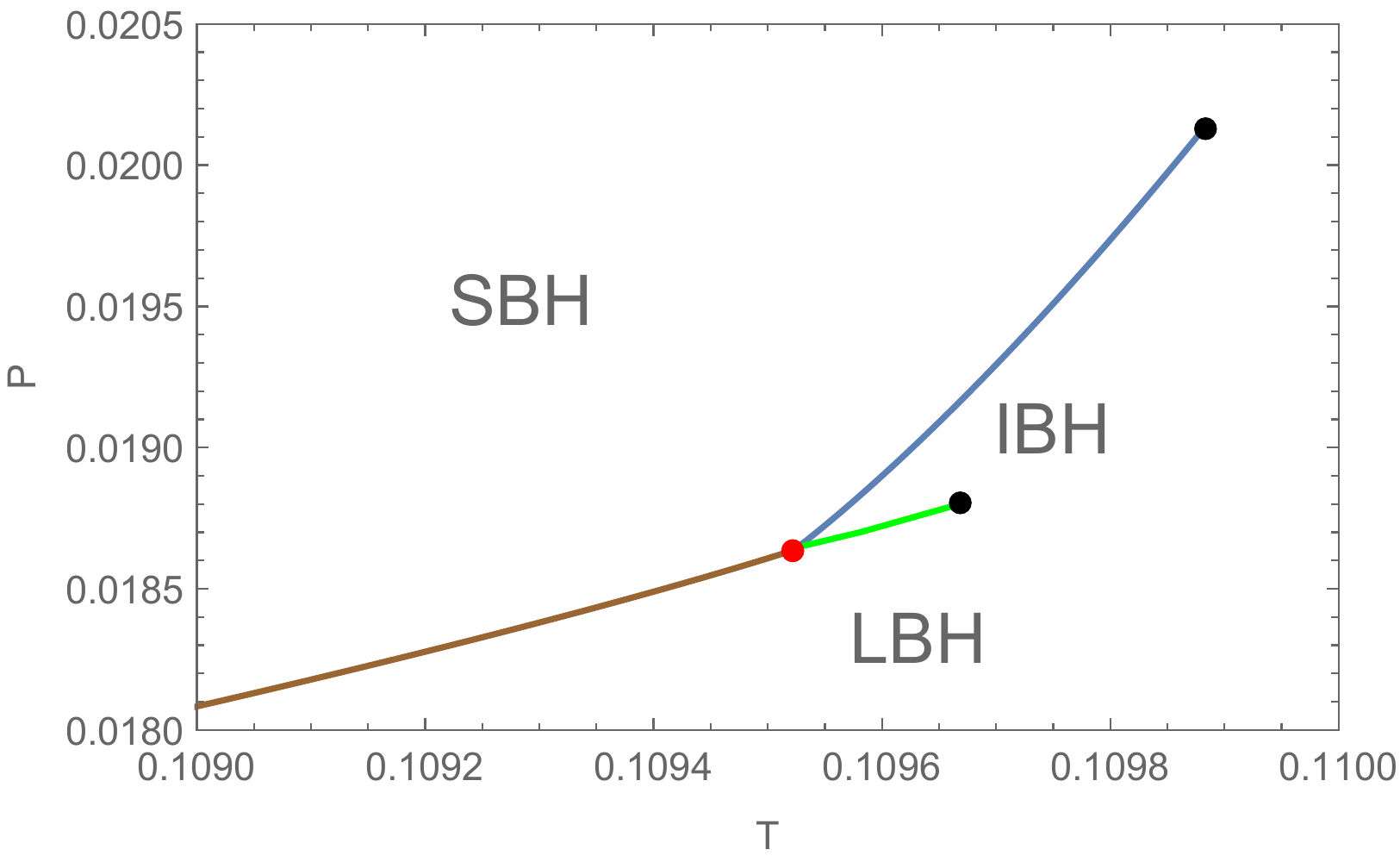}\\
  \caption{Phase diagram for $D=6$ with the triple point. In this plot, $q=0.08$ and $\alpha=1.05$. The black point represents the critical point. "SBH", "IBH" and "LBH" represent the small, the intermediate, and the large Gauss-Bonnet black holes respectively.}
  \label{Phase_diagram_d6_three}
\end{figure}

Now, we consider the case that the system has a triple point on the phase diagram. In Fig.\ref{Phase_diagram_d6_three}, the phase diagram for $q=0.08$ and $\alpha=1.05$ is plotted. There are two critical points. One critical point with the coordinates $(0.109883, 0.0201336)$ denotes the end point of the coexisting curve for the small black hole phase and the intermediate black hole phase, and the other one with the coordinates $(0.109669, 0.0188013)$ is the end point of the coexisting curve for the large black hole and the intermediate black hole. In addition, there is a triple point with the coordinates $(0.109523, 0.0186373)$, denoting the coexisting phase of the three branches of black holes.

\begin{figure}
  \centering
  \includegraphics[width=8cm]{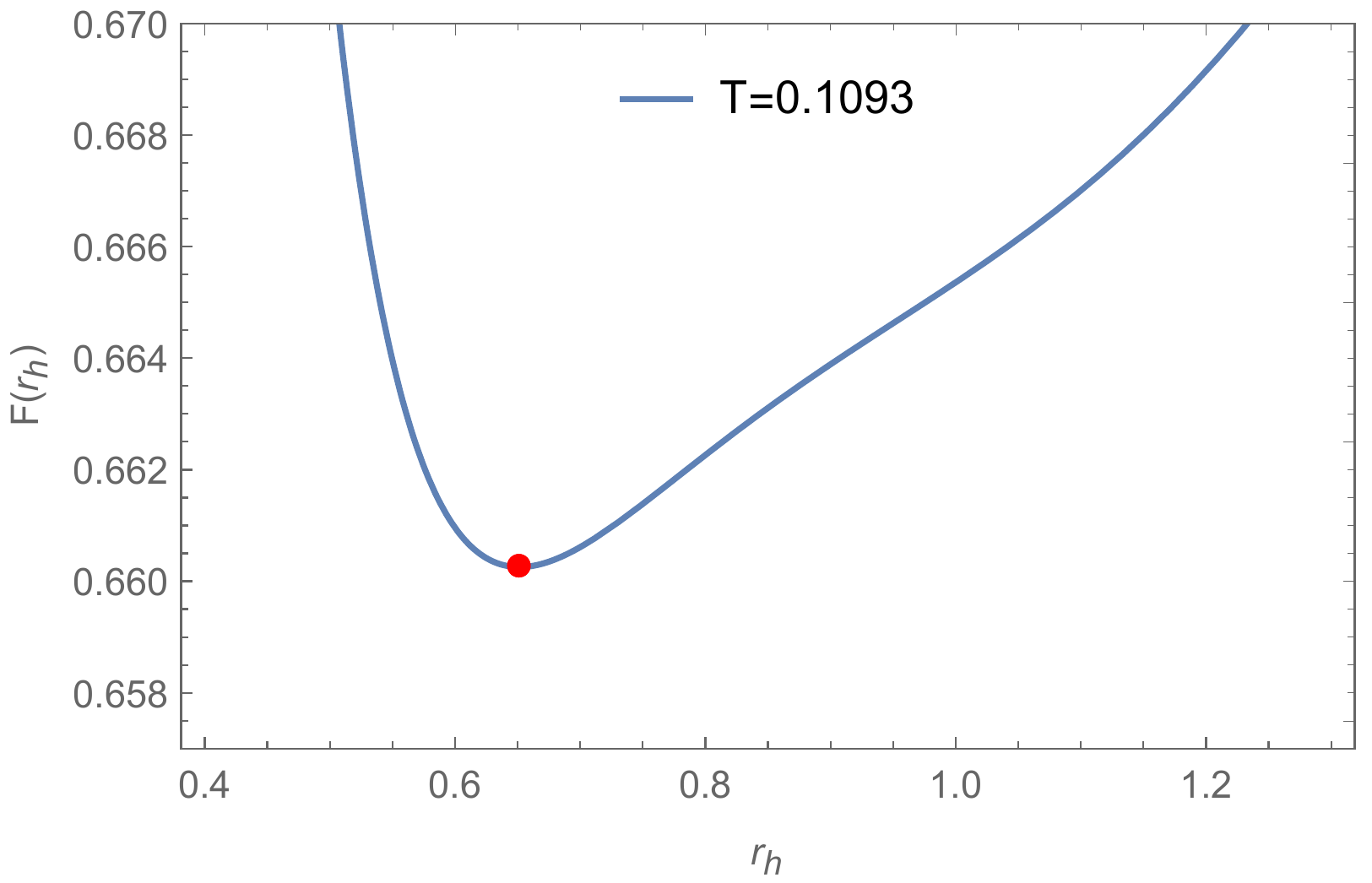}
  \includegraphics[width=8cm]{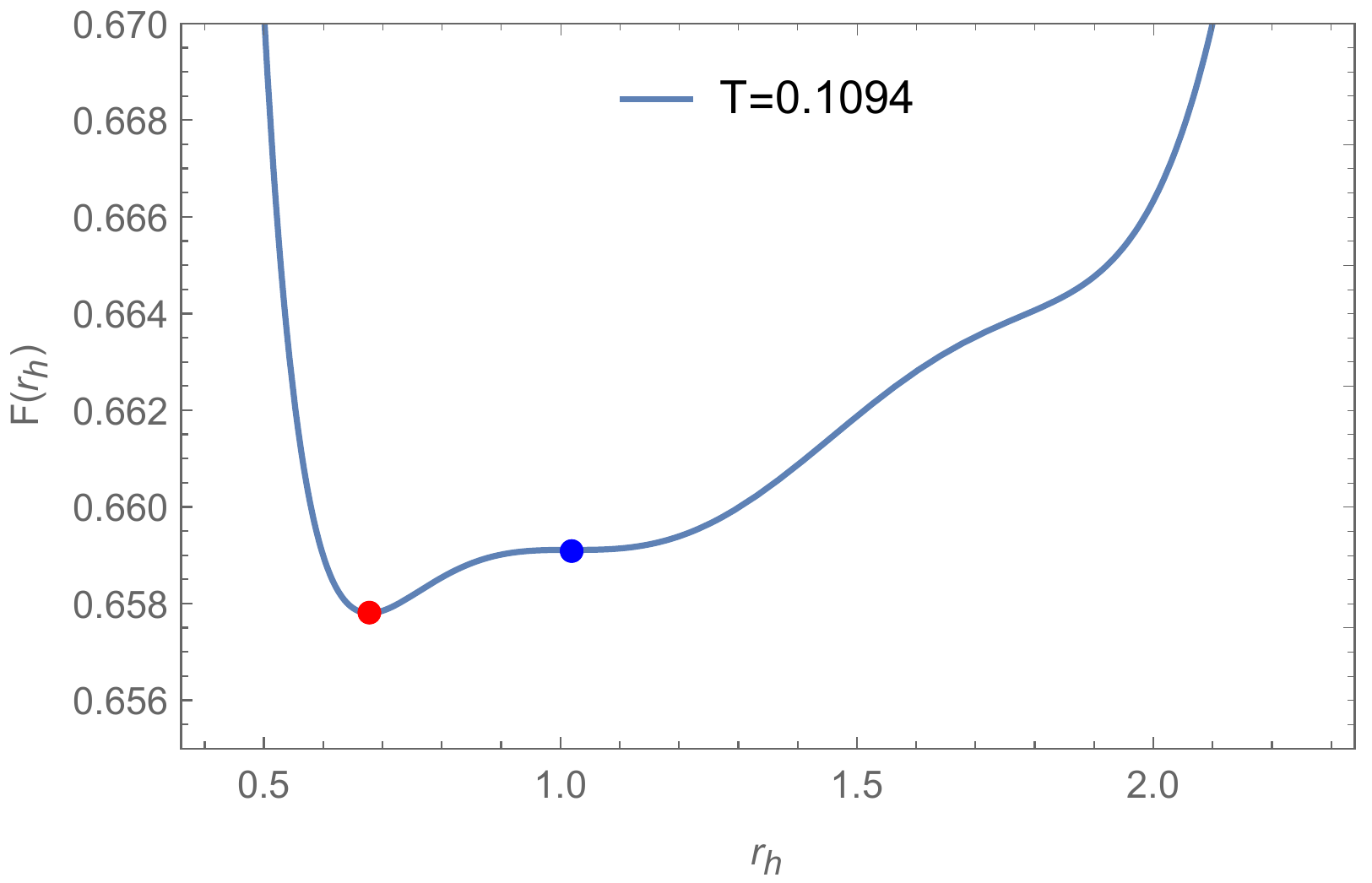}\\
  \includegraphics[width=8cm]{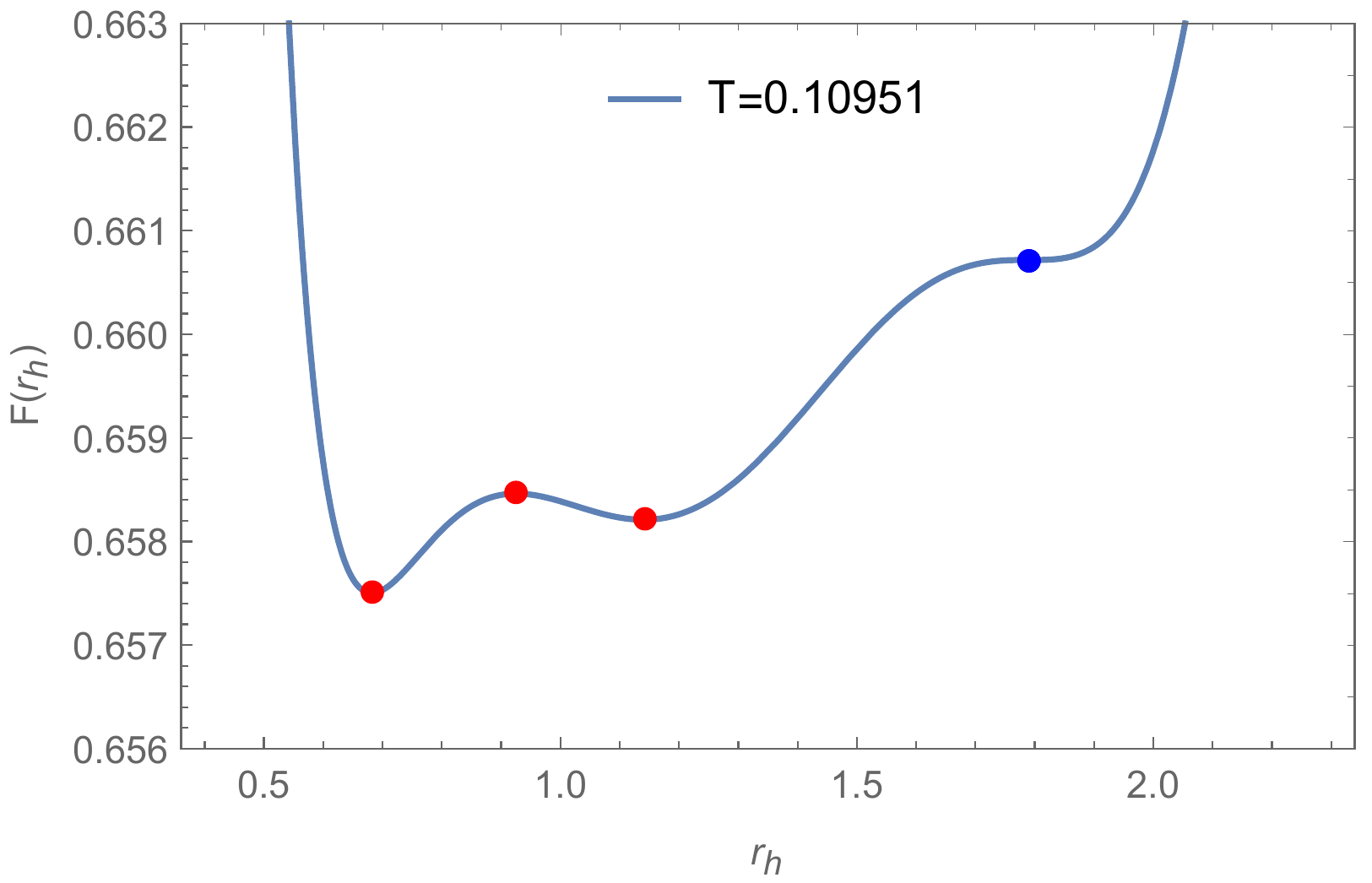}
  \includegraphics[width=8cm]{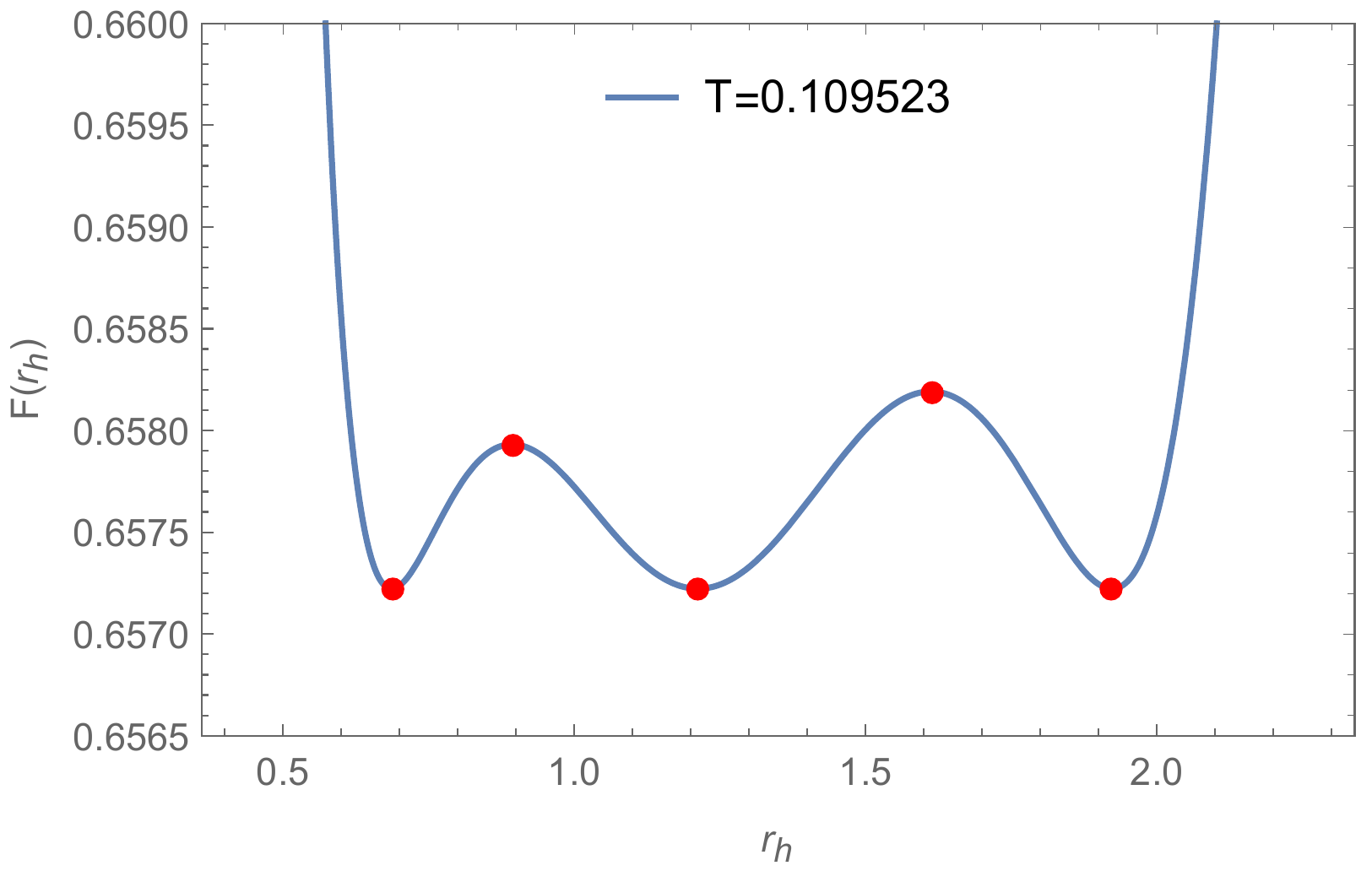}\\
  \includegraphics[width=8cm]{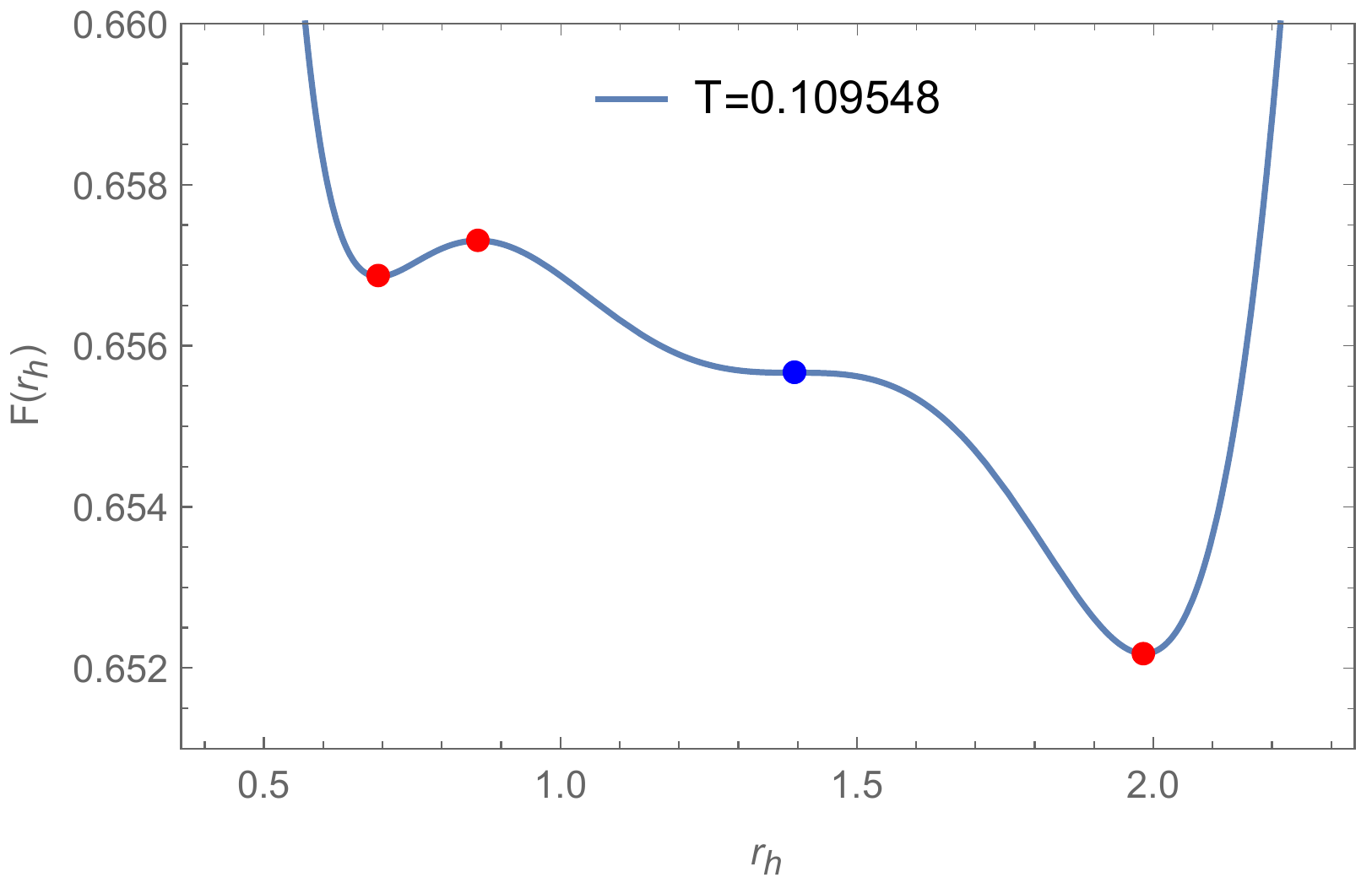}
  \includegraphics[width=8cm]{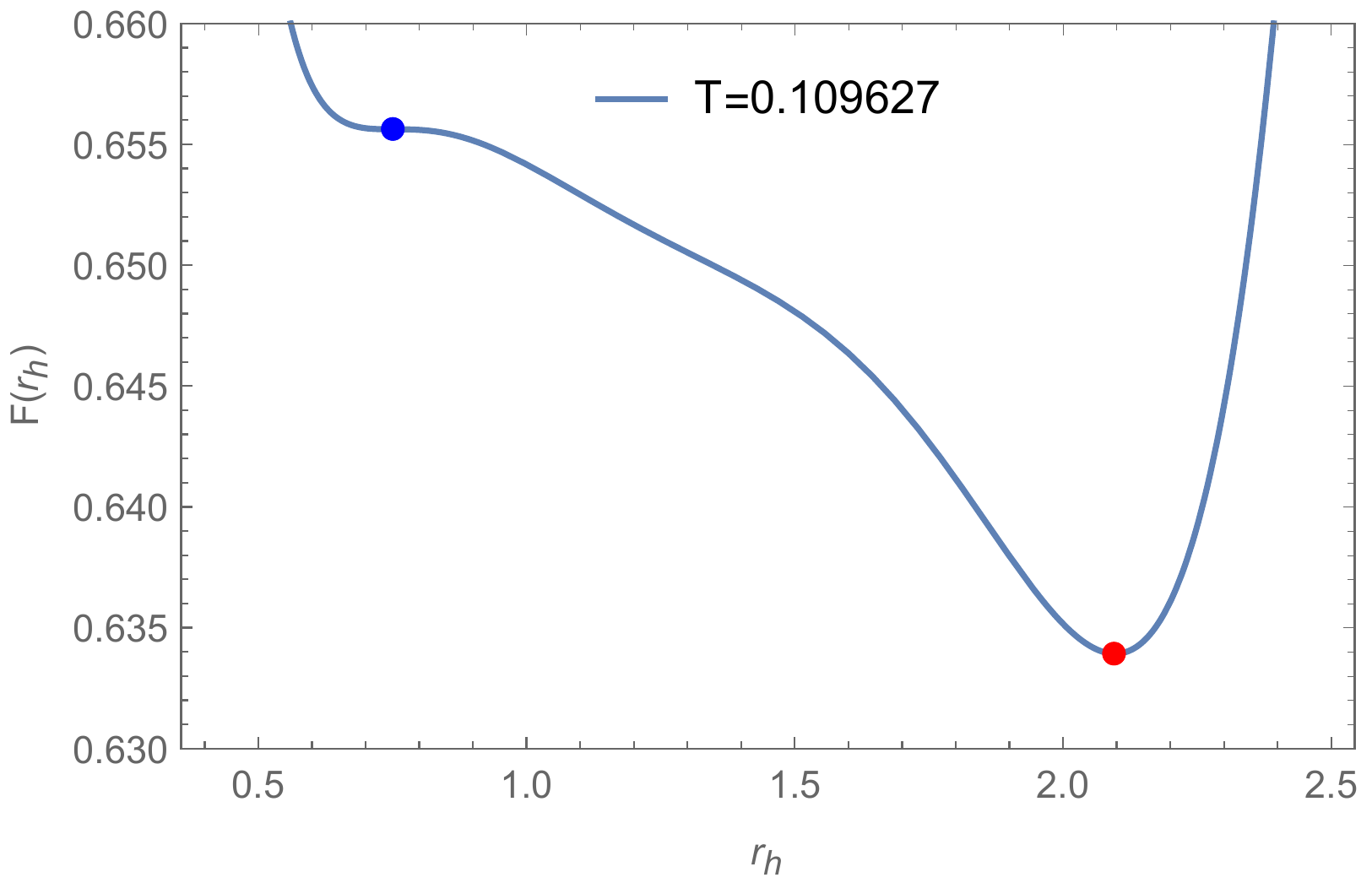}\\
  \caption{Free energy landscape for $D=6$. In this plot, $P=0.0186373$, $\alpha=1.05$, and $q=0.08$. The red points represent the equilibrium state Gauss-Bonnet black holes, while the blue points represent the inflection points on the landscape. }
  \label{Landscape_d6_three}
\end{figure}

We also plot the free energy landscapes for this case in Fig.\ref{Landscape_d6_three}. When $T<0.1094$, only one stable black hole phase with the relatively small radius appears on the landscape. As the temperature increases to $0.1094$, an inflection point appears, which means that another locally stable black hole phase with the intermediate radius appears. As the temperature increases to $0.10951$, the second inflection point appears, which means that the third locally stable black hole phase emerges. At the temperature $T=0.109523$, the three locally stable black hole phases coexist, with the same depths of the potential well. When we continue to increase the temperature, the third and the forth inflection points will appear successively, which implies that the intermediate and the small black hole phases disappear correspondingly. At last, when $T>0.109627$, there is left only one stable black hole phase with the relatively large radius on the landscape.

\subsection{$D\ge 7$}

We state that the $D\ge7$ case is similar to the $D=5$ case. There is only one critical point on the phase diagram, which is the endpoint of the coexisting curve of the small and the large Gauss-Bonnet black holes. When the pressure is lower than the critical pressure, increasing the ensemble temperature, the free energy landscape will change from single well to double well and restore the shape of single well. When the temperature is lower/higher than transition temperature, the single potential well is located at the small/large black hole radius, which means the small/large black hole state is thermodynamically stable.

\section{Generalized free energy in grand canonical ensemble} 
\label{SecV}

In this section, we discuss the generalized free energy for Gauss-Bonnet gravity in grand canonical ensemble. In Sec.\ref{SecIII}, we have derived the generalized free energy function in the canonical ensemble by using the Euclidean path integral approach. The generalized free energy in grand canonical ensemble can be obtained by using Legendre transformation  
\begin{eqnarray}
\Omega&=&F-Q\Phi
\nonumber\\
&=&\frac{1}{16\pi} (D-2) \Omega_{D-2} \left(r_h^{D-3}+\frac{16\pi}{(D-1)(D-2)}P r_h^{D-1}+\alpha r_h^{D-5}\right)+\frac{Q^2}{2(D-3)\Omega_{D-2} r_h^{D-3}}\nonumber\\
&&-\frac{1}{4} T \Omega_{D-2} r_h^{D-2} \left(1+\frac{2\alpha(D-2)}{(D-4)}\frac{1}{r_h^2}\right)-Q\Phi\;,
\end{eqnarray}
where we have replaced the parameter $q$ with the electric charge $Q$ by using the relation in Eq.(\ref{M_Q_S}). The electromagnetic potential $\Phi$ should be viewed as the thermodynamic potential of the bath in the grand canonical ensemble. It should be noted that the generalized free energy $\Omega$ is the function of the order parameter $r_h$ and $q$ while $T$ and $\Phi$ are the external adjustable parameters.   

\begin{figure}
  \centering
  \includegraphics[width=8cm]{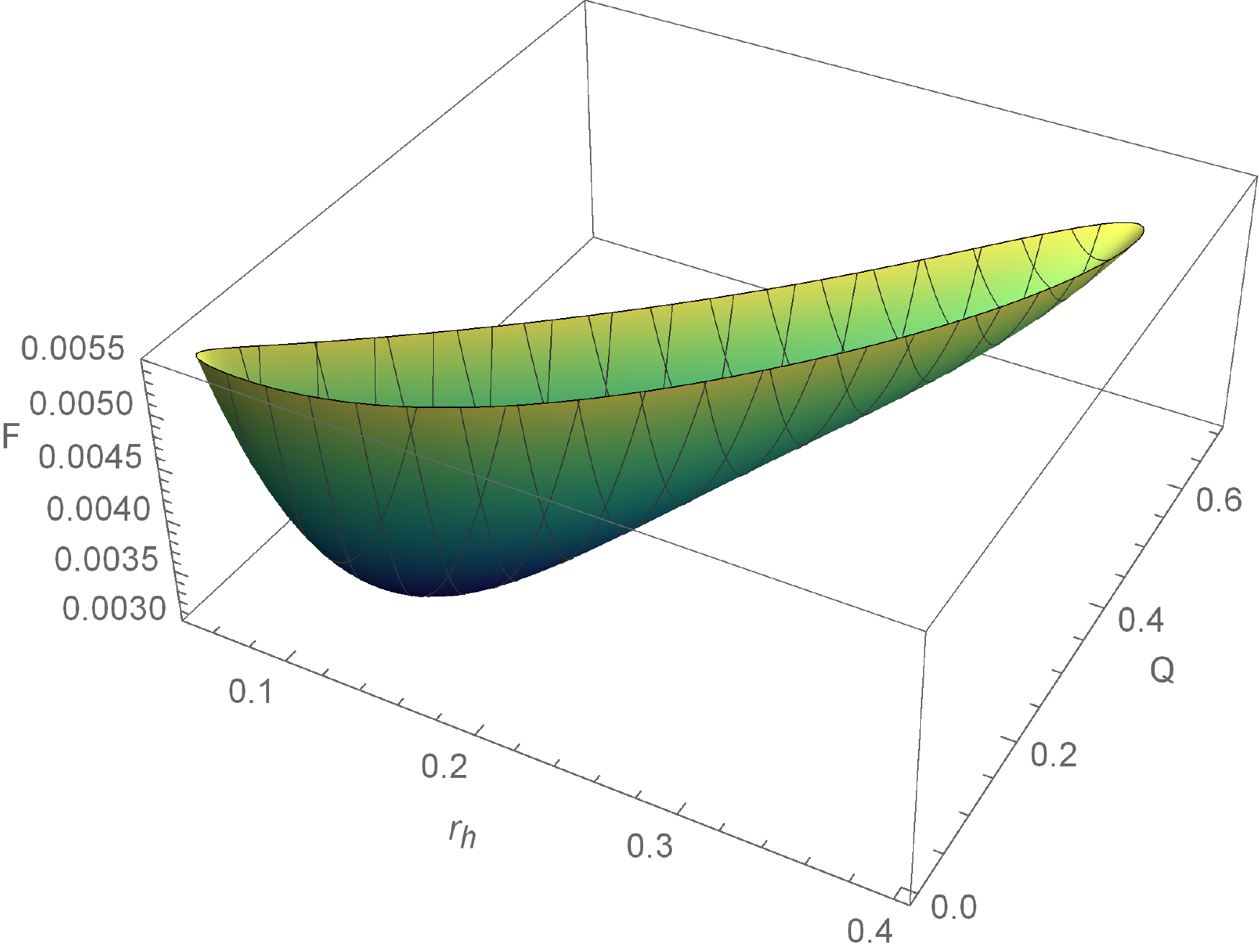}
  \includegraphics[width=8cm]{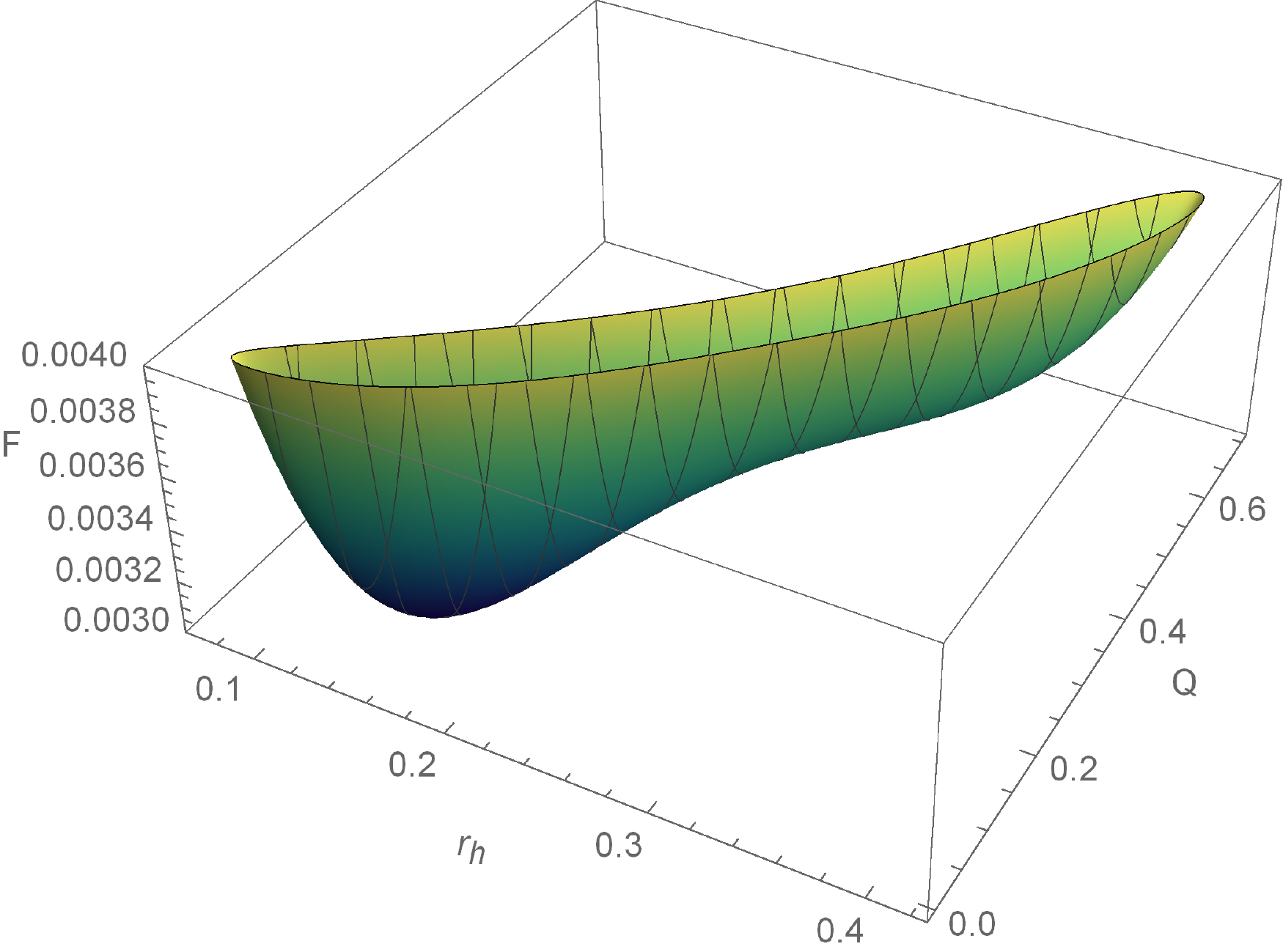}\\
  \includegraphics[width=8cm]{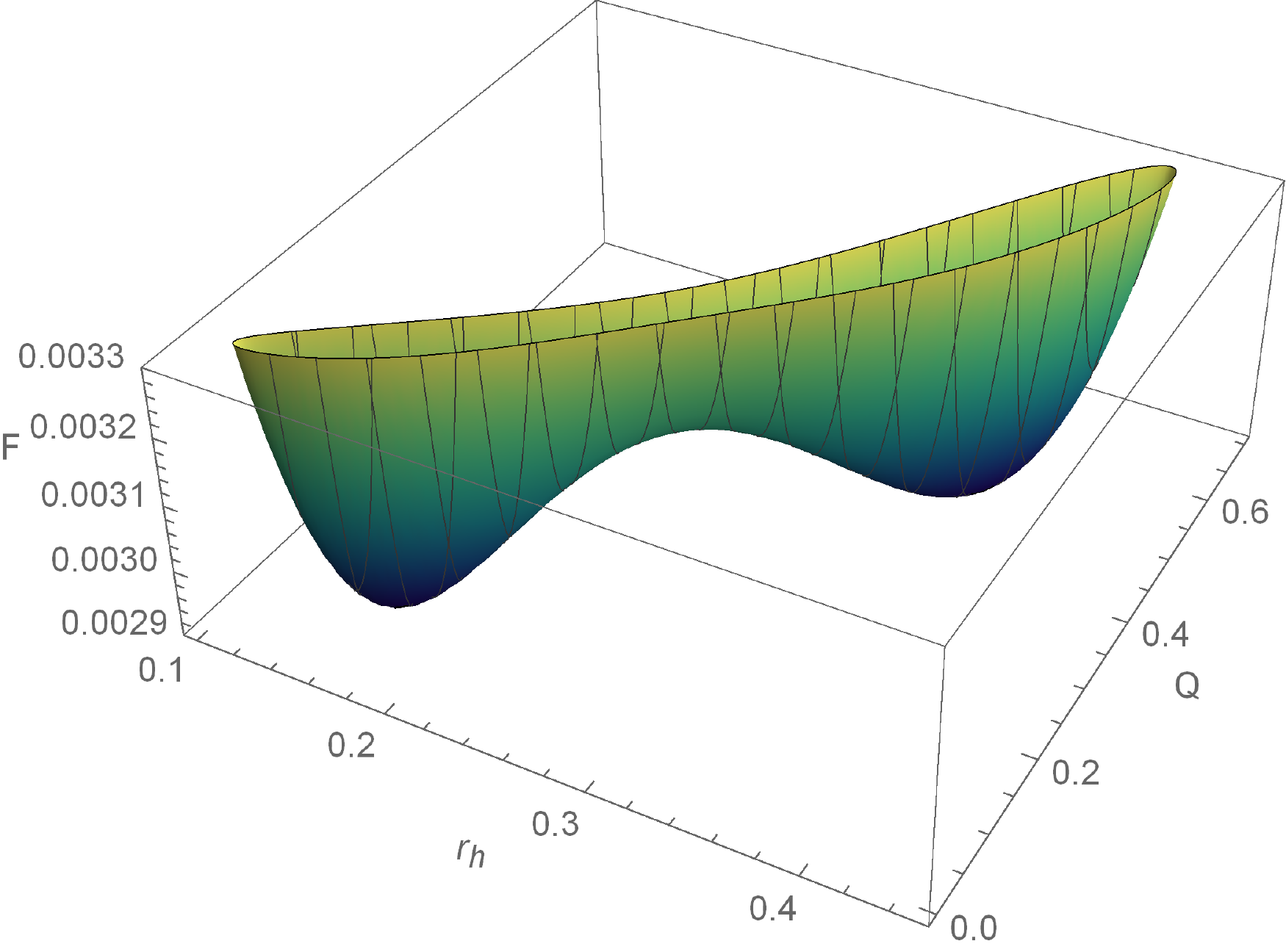}
  \includegraphics[width=8cm]{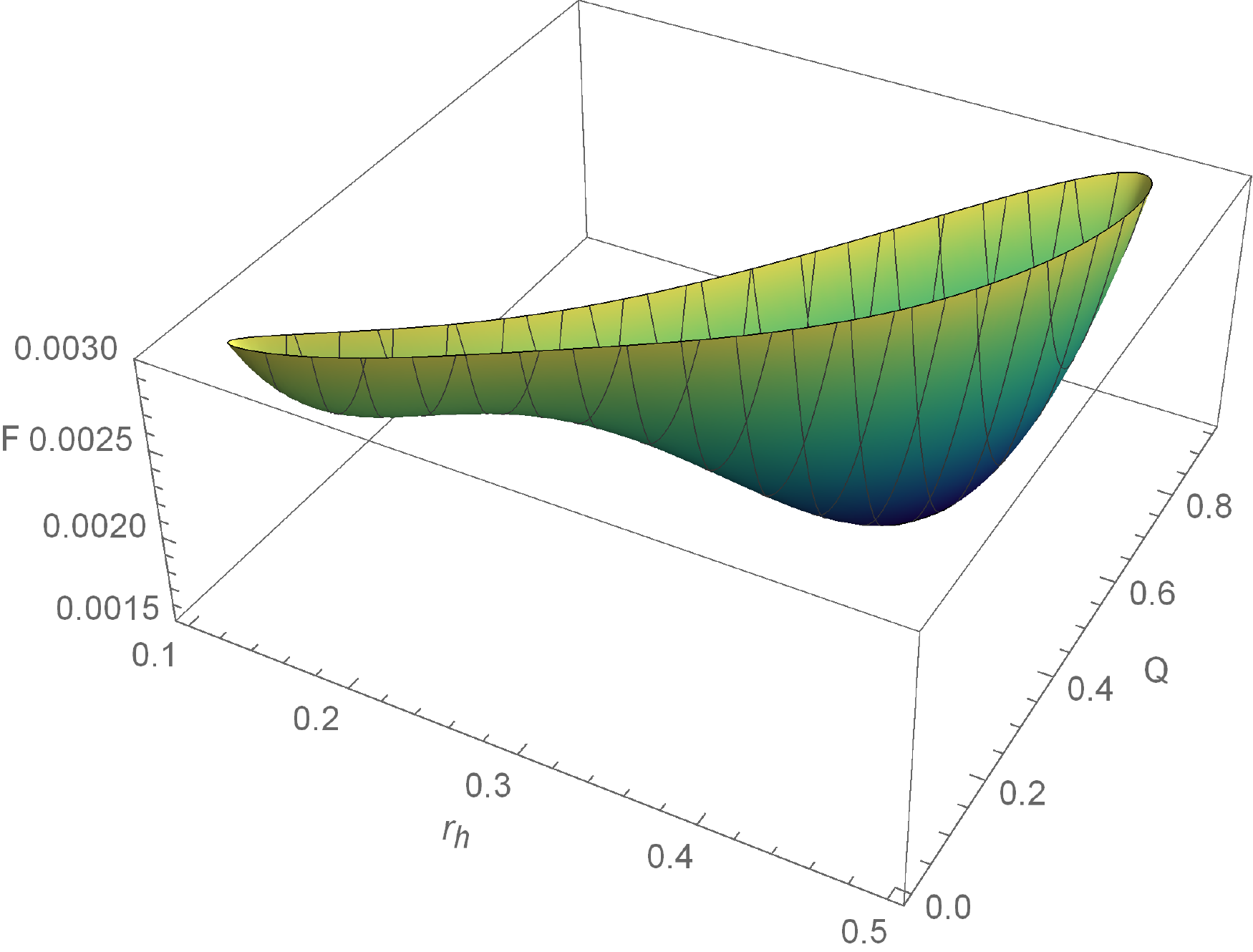}\\
  \caption{Two dimensional free energy landscape for $D=5$. In this plot, $P=0.5$, $\alpha=0.01$, and $\Phi=0.1$. The ensemble temperatures are set to be $0.52$, $0.524813$, $0.526749$, and $0.530519$ respectively. }
  \label{2d_Landscape}
\end{figure}

In \cite{Cai:2013qga,Zou:2014mha}, it was shown that there exists the small/large Gauss-Bonnet black hole phase transition in five dimensions. For $D=5$, we have 
\begin{eqnarray}
\Omega=F-Q\Phi&=&\frac{3\pi}{8}\left(r_h^2+\frac{4\pi}{3}P r_h^{4}+\alpha +\frac{Q^2}{3\pi^3 r_h^2}\right)-\frac{\pi^2}{2} T r_h^{3} \left(1+\frac{6\alpha}{r_h^2}\right)-Q\Phi\;.
\end{eqnarray}
This is just the generalized free energy that was defined previously by using the thermodynamic relation in \cite{Li:2023ppc}. In order to describe the thermodynamics of the phase transition qualitatively, we plot the corresponding free energy landscapes of the grand canonical ensemble at different ensemble temperatures in Fig.\ref{2d_Landscape}.

In Fig.\ref{2d_Landscape}, we set $p=0.5$, $\alpha=0.01$, and $\Phi=0.1$. It is shown that when $T<0.524813$ and $T>0.530519$, the shape of free energy landscapes is a single well, and when $0.524813<T<0.530519$, the shape is a double well. AT the phase transition point, $T=0.526749$, the corresponding landscape has two potential wells with the same depths. All these behavior are similar to that of landscapes for canonical ensemble in $D=5$.

\section{conclusion and discussion} 
\label{SecVI}

In summary, we have derived the generalized free energy function of the $D$-dimensional charged Gauss-Bonnet AdS black holes in terms of the path integral approach. It is demonstrated that the derived generalized free energy is consistent with the thermodynamic definition. We also discuss the free energy landscapes for the Gauss-Bonnet gravity in diverse dimensions. For the canonical ensemble, the free energy landscapes are one dimensional curves with the stable states represented by the lowest points in the potential wells. Based on the generalized free energy function and the landscapes, we discuss the corresponding phase structures that are illustrated on the phase diagrams. For $D=5$ and $D\ge7$, there is only one critical point on the phase diagram. For $D=6$, there exists two cases: one case with only one critical point on the phase diagram and another case with two critical points and one triple point on the phase diagram. We have explicitly plotted the landscapes to exhibit the changes with the ensemble temperature. In addition, based on the topography of the landscape, we discussed the thermodynamics of the state switching and the phase transition. For the Gauss-Bonnet black holes in the grand canonical ensemble, the landscapes are two dimensional surfaces. We briefly study the shapes of these landscapes at different temperatures.

In the present work, the thermodynamics of the state switching and the phase transition for the charged Gauss-Bonnet AdS black holes is discussed in detail. However, the thermodynamics alone cannot provide the full information for the kinetics of the state switching and the phase transition. The thermal fluctuations should be taken into account in order to investigate the phase transition kinetics. For future directions, it is interesting to study the state switching and the phase transition process on the two dimensional landscapes by using the stochastic dynamics method.

\end{document}